\documentclass[]{pasj00}

\begin{document}
\SetRunningHead{Author(s) in page-head}{Running Head}
\Received{****/**/**}
\Accepted{****/**/**}

\title{A Suzaku Study of Ejecta Structure and Origin of Hard X-ray Emission in the Supernova Remnant G156.2+5.7}

\author{Hiroyuki \textsc{Uchida}\altaffilmark{1}, Hiroshi \textsc{Tsunemi}\altaffilmark{2}, Satoru \textsc{Katsuda}\altaffilmark{3}, Koji \textsc{Mori}\altaffilmark{4}, Robert \textsc{Petre}\altaffilmark{5}, and Hiroya \textsc{Yamaguchi}\altaffilmark{6}} %
\altaffiltext{1}{Department of Physics, Graduate School of Science, Kyoto University, Kitashirakawa Oiwake-cho, Sakyo-ku, Kyoto 606-8502, Japan}
\altaffiltext{2}{Department of Earth and Space Science, Graduate School of Science, Osaka University, Toyonaka, Osaka 560-0043, Japan}
\altaffiltext{3}{RIKEN, 2-1 Hirosawa, Wako, Saitama 351-0198, Japan}
\altaffiltext{4}{Department of Applied Physics, Faculty of Engineering, University of Miyazaki, 1-1 Gakuen Kibana-dai Nishi, Miyazaki 889-2192, Japan}
\altaffiltext{5}{Code 662, NASA Goddard Space Flight Center, Greenbelt, MD 20771, USA}
\altaffiltext{6}{Harvard-Smithsonian Center for Astrophysics, 60 Garden St., Cambridge, MA 02138, USA}
\email{uchida@cr.scphys.kyoto-u.ac.jp}


\maketitle

\begin{abstract}
We report an X-ray study of the evolved Galactic supernova remnant (SNR) G156.2+5.7 based on six pointing observations with Suzaku.
The remnant's large extent (100$\arcmin$ in diameter) allows us to investigate its radial structure in the northwestern and eastern directions from the apparent center.
The X-ray spectra were well fit with a two-component non-equilibrium ionization model representing the swept-up interstellar medium (ISM) and the metal-rich ejecta.
We found prominent central concentrations of Si, S and Fe from the ejecta component; the lighter elements of O, Ne and Mg were distributed more uniformly.
The temperature of the ISM component suggests a slow shock (610-960 km s$^{-1}$), hence the remnant's age is estimated to be 7,000-15,000 yr, assuming its distance to be $\sim$1.1 kpc.
G156.2+5.7 has also been thought to emit hard, non-thermal X-rays, despite being considerably older than any other such remnant. 
In response to a recent discovery of a background cluster of galaxies (2XMM J045637.2+522411), we carefully excluded its contribution, and reexamined the origin of the hard X-ray emission.
We found that the residual hard X-ray emission is consistent with the expected level of the cosmic X-ray background.
Thus, no robust evidence for the non-thermal emission was obtained from G156.2+5.7.
These results are consistent with the picture of an evolved SNR.

\end{abstract}

\section{Introduction}
G156.2+5.7 (RX04591+5147) was initially discovered in X-ray with ROSAT \citep{Pfeffermann91}. 
Its age was estimated at $\sim$26000 yr, based on the Sedov model, hence is thought to be an evolved supernova remnant (SNR).
The distance is comparatively nearby ($\sim$1.1 kpc; \cite{Pfeffermann91}) and its large apparent size ($\sim$100\arcmin \ in diameter) enables us to study the detailed plasma structure.
While G156.2+5.7 is one of the brightest SNRs in X-rays, its radio emission is quite faint; this is its most remarkable feature.
Shortly after its discovery in X-rays, the radio continuum emission was also detected with the Effelsberg 100-m telescope \citep{Reich92}.
As they point out, the surface brightness at 1 GHz (5.8$\times$10$^{-23}$ W m$^{-2}$ Hz$^{-1}$ sr$^{-1}$) is the lowest among all the known SNRs.
The radio morphology along the northwest and southeast rim shows limb brightening (\cite{Reich92}; \cite{Xu07}) that is typically seen in so-called "barrel-shaped" SNRs \citep{Kesteven87}.
In contrast to its radio morphology, G156.2+5.7 has centrally-filled thermal emission in X-rays.
While \citet{Lazendic06} lists G156.2+5.7 as a mixed-morphology (MM) SNR for this reason, it does not fit the definition of MM SNRs in several ways:  the X-ray shell is easily distinguishable and there are neither CO clouds nor OH (1720 MHz) masers in its vicinity. 
A recent optical observation suggests that the distance to G156.2+5.7 is smaller than that previously estimated \citep{Gerardy07}.
Their main argument is that several bright H$\alpha$ filaments correlate with nearby interstellar clouds ($\sim$0.3 kpc).
However, the X-ray and the radio observations both agree that G156.2+5.7 is located at 1-3 kpc, i.e., an evolved SNR (\cite{Pfeffermann91}; \cite{Reich92}).
The distance and the age of this SNR remain open questions.
\begin{figure*}[!t]
  \begin{center}
    \FigureFile(100mm,100mm){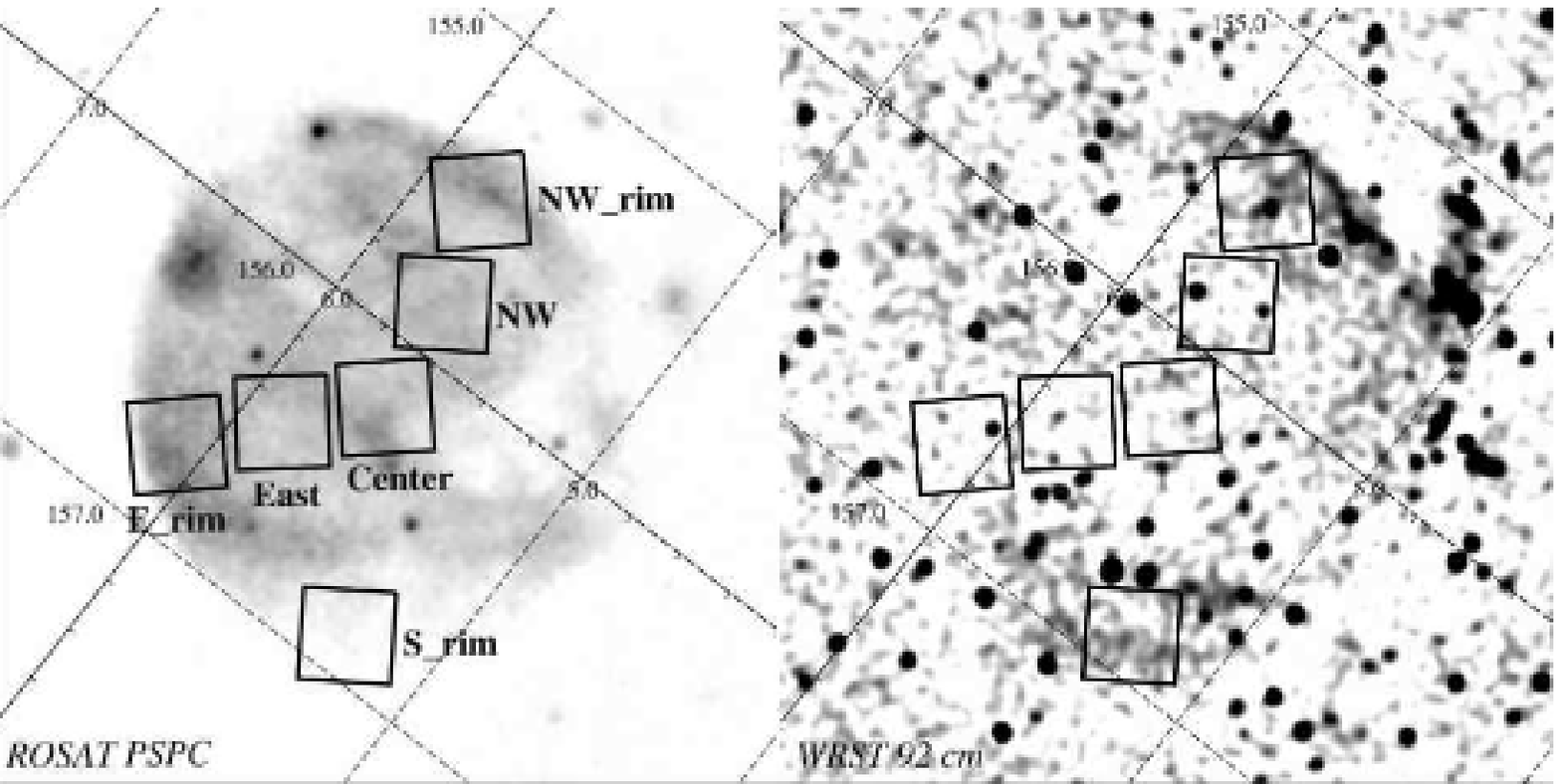}
    \FigureFile(60mm,60mm){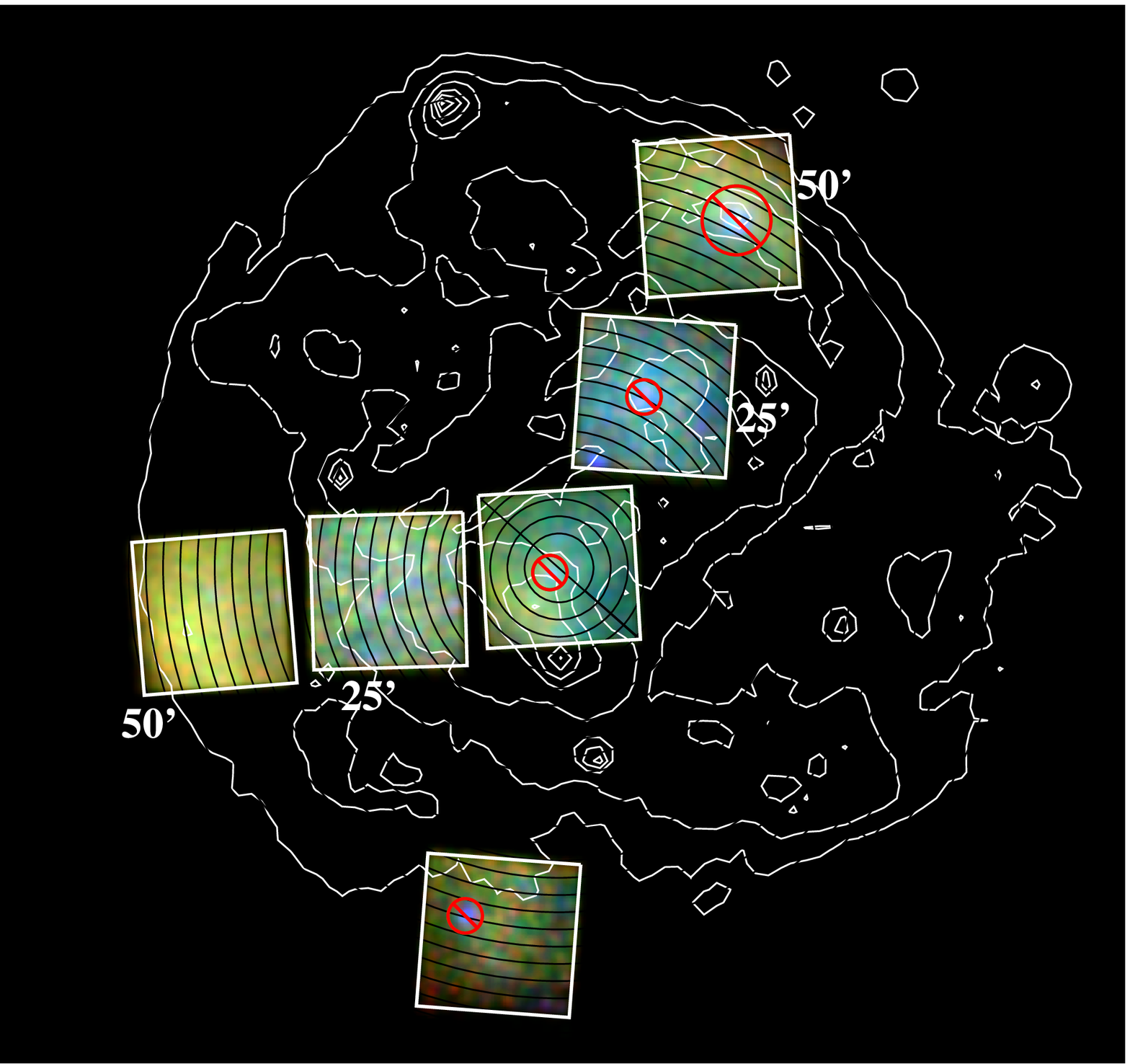}
  \end{center}
  \caption{\textit{Left}: ROSAT PSPC image of G156.2+5.7 with the Galactic coordinate overlaid. The FoV of \textit{Suzaku} XIS are shown with solid squares. \textit{Middle}: Radio (92 cm) image of G156.2+5.7 obtained by the Westerbork Northern Sky Survey (WENSS) with the Westerbork Synthesis Radio Telescope (WSRT). The coordinate scales are matched in both images. \textit{Right}: True-color XIS image of G156.2+5.7. The energy bands of red, green and blue correspond to 0.5-0.7 keV, 0.7-1.0 keV and 1.0-5.0 keV, respectively. The red circular regions were excluded from our spectral analyses.}\label{fig:image}
\end{figure*}
\begin{table*}[!t]
\caption{Summary of the 6 \textit{Suzaku} observations\label{tab:obs}}
\begin{center}
\begin{tabular}{lcccc}
\hline
\hline
Name & Obs. ID  & Obs. Date & R.A., Dec. (J2000) & Effective Exposure\\
\hline
	   						    		      			  						
NW\_rim  & 501075010   & 2007-Feb-16 &  04$^{\mathrm h}$56$^{\mathrm m}$54$^{\mathrm s}$.9, 52\arcdeg24\arcmin12\arcsec.6  & 50.5 ks\\ 
	   					    		      			  		    			
Center & 501106010   & 2007-Feb-17 &  04$^{\mathrm h}$58$^{\mathrm m}$54$^{\mathrm s}$.9, 51\arcdeg44\arcmin36\arcsec.1  & 51.2 ks\\ 

E\_rim   & 501074010   & 2007-Feb-18 &  05$^{\mathrm h}$03$^{\mathrm m}$13$^{\mathrm s}$.5, 51\arcdeg37\arcmin06\arcsec.2  & 53.3 ks\\ 
							    		      			  		    			
East   & 504082010   & 2010-Feb-21 &  05$^{\mathrm h}$01$^{\mathrm m}$03$^{\mathrm s}$.0, 51\arcdeg41\arcmin41\arcsec.6  & 50.3 ks\\ 
	   						    						    		    
NW     & 504081010   & 2010-Mar-03 &  04$^{\mathrm h}$57$^{\mathrm m}$42$^{\mathrm s}$.1, 52\arcdeg04\arcmin31\arcsec.1  & 52.9 ks\\ 
	   						    						    		    
S\_rim   & 504080010   & 2010-Mar-04 &  04$^{\mathrm h}$59$^{\mathrm m}$41$^{\mathrm s}$.8, 51\arcdeg00\arcmin36\arcsec.0  & 52.6 ks\\ 
	   						    						    		    
\hline
\end{tabular}
\end{center}
\end{table*}

The X-ray spectrum of G156.2+5.7 is also largely typical of an evolved SNR.
\citet{Yamauchi93} observed G156.2+5.7 with the Ginga LAC (large area proportional counters) and detected extended X-ray emission in the 1.2-3.0 keV energy band.
\citet{Yamauchi99} performed an additional observation with ASCA and found that its spectrum consists of two components with different features.
The soft component has several emission lines, suggesting a thin thermal origin whereas the hard component is featureless. 
However, due to the lack of statistics, they were not able to decide whether the featureless hard component has a thermal origin or a non-thermal origin.
\citet{Katsuda09} observed two rims (northwest and east) and the center with Suzaku.  
They found two thermal components, one of which they ascribed to interstellar material heated by the forward shock, and the other reverse-shock heated ejecta.  
They also detected hard X-ray observed from the northwestern rim and the center, which they concluded is likely to be non-thermal synchrotron emission.
If this is the case, G156.2+5.7 would be the oldest SNR producing non-thermal X-ray emission. 
However, a recent Suzaku observation discovered a cluster of galaxies (2XMM J045637.2+522411; \cite{Yamauchi10}) located in the northwestern rim.
Since the surface brightness of this extended source is higher than that of the rim, we need careful consideration of its contribution to determine the true origin of the hard component.

\citet{Katsuda09} also estimated the progenitor mass to be less than 15\MO \ from the metal abundances of the thermal emission detected in the central portion.
While no compact source has been discovered in G156.2+5.7 (see \cite{Kaplan06}), their result suggests that its origin is a core-collapse supernova (SN) rather than a type Ia SN.
More recently, \citet{Hudaverdi10} observed the center of this remnant with XMM-Newton and measured its metallicity. 
From the results, they suggested that G156.2+5.7 is an O-rich SNR.
 
We performed 6 pointing observations of G156.2+5.7 with Suzaku.
The fields of view (FoV) cover three sections of the rim and three inner regions, and include the ones discussed in \citet{Katsuda09}.
In this paper, we report the ejecta distribution and review the origin of the hard X-ray emission detected by \citet{Katsuda09} along with the newly observed regions.
We also discuss the age of this SNR based on these analyses.

\section{Observations and Data Reduction}
We summarize the 6 observations in table \ref{tab:obs}. 
Data were taken in 2007 and 2010, during the Suzaku AO1 and AO4 phases, respectively.
The observed regions are shown in figure \ref{fig:image}. 
The squares represent the FoV of the Suzaku XIS (X-ray Imaging Spectrometer; \cite{Koyama07}).
The data were analyzed with version 6.9 of the HEAsoft tools. 
For data reduction we used version 9 of the Suzaku Software. 
The calibration database (CALDB) used was the one updated in March 2010. 
We used revision 2.2 of the cleaned event data and combined the 3$\times$3 and 5$\times$5 event files. 
In order to exclude background flare events, we obtained the good time intervals (GTIs) by including only times at which the count rates in the various detectors are less than $+3\sigma$ of the mean count rates.

\begin{table*}[!tb]
\caption{Surface brightnesses of hard emission in G156.2+5.6 and two nearby observations\label{tab:cxb}}
\begin{center}
\begin{tabular}{lccc}
\hline
\hline
Name  & $l$ & $b$ & surface brightness at 2.0-10.0 keV\\
& [deg] & [deg] & [10$^{-11}$ erg cm$^{-2}$s$^{-1}$deg$^{-2}$]\\
\hline
G156.2+5.7 NW\_rim &   155.5  & 5.8 & 1.88$\pm$0.26 \\
G156.2+5.7 E\_rim &  156.7  & 6.1 & 1.61$\pm$0.23\\ 
G156.2+5.7 S\_rim &   156.8  & 5.3 & 1.56$\pm$0.22\\
IRAS 05262+4432 BGD & 165.0  & 5.6 &  1.78$\pm$0.25\\
SGR 0501+4516 BGD & 161.4 & 1.9 &  1.89$\pm$0.26\\
\hline
\end{tabular}
\end{center}
\end{table*}
\begin{figure*}[!t]
  \begin{center}
    \FigureFile(70mm,70mm){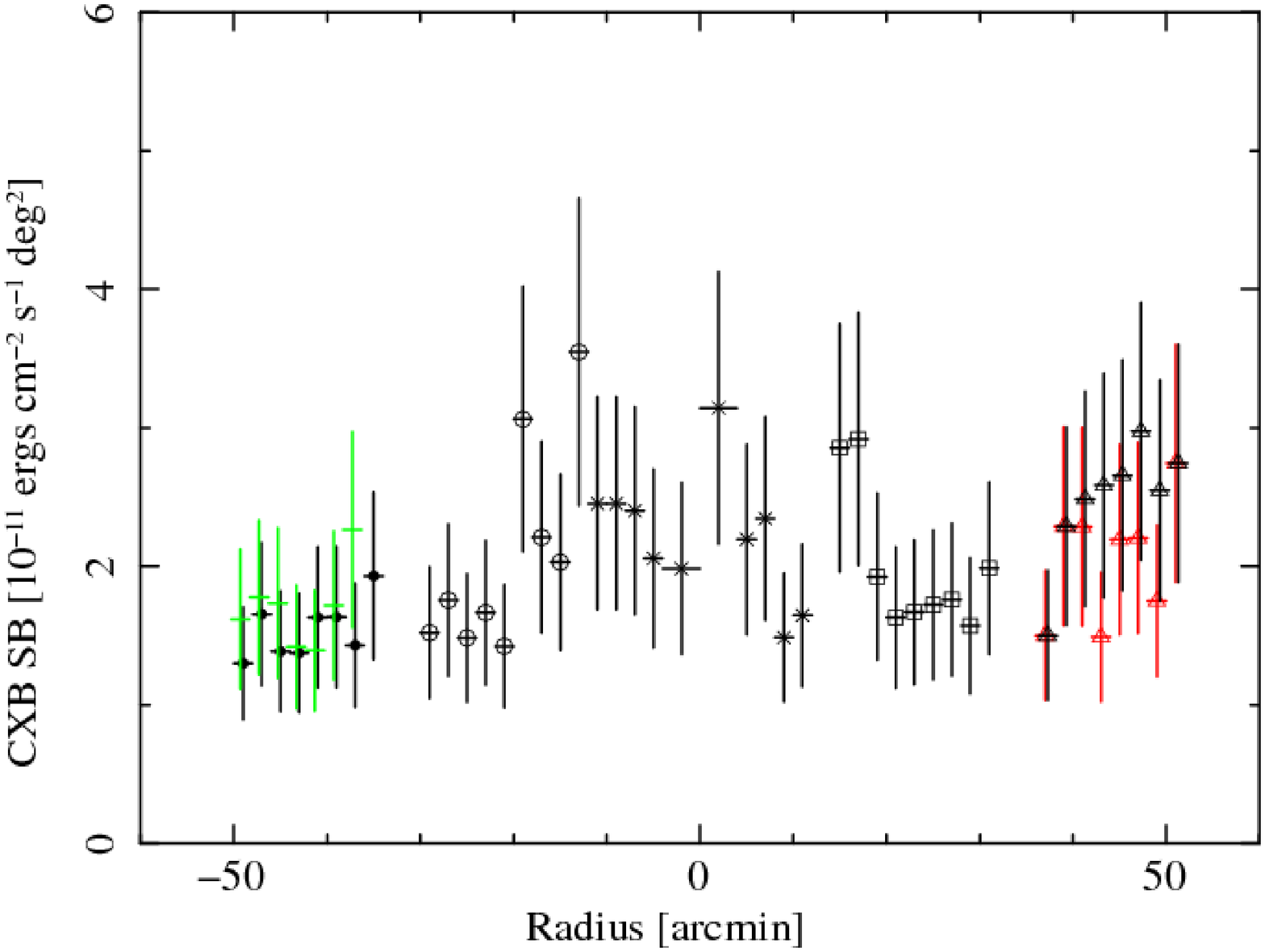}
    \FigureFile(70mm,70mm){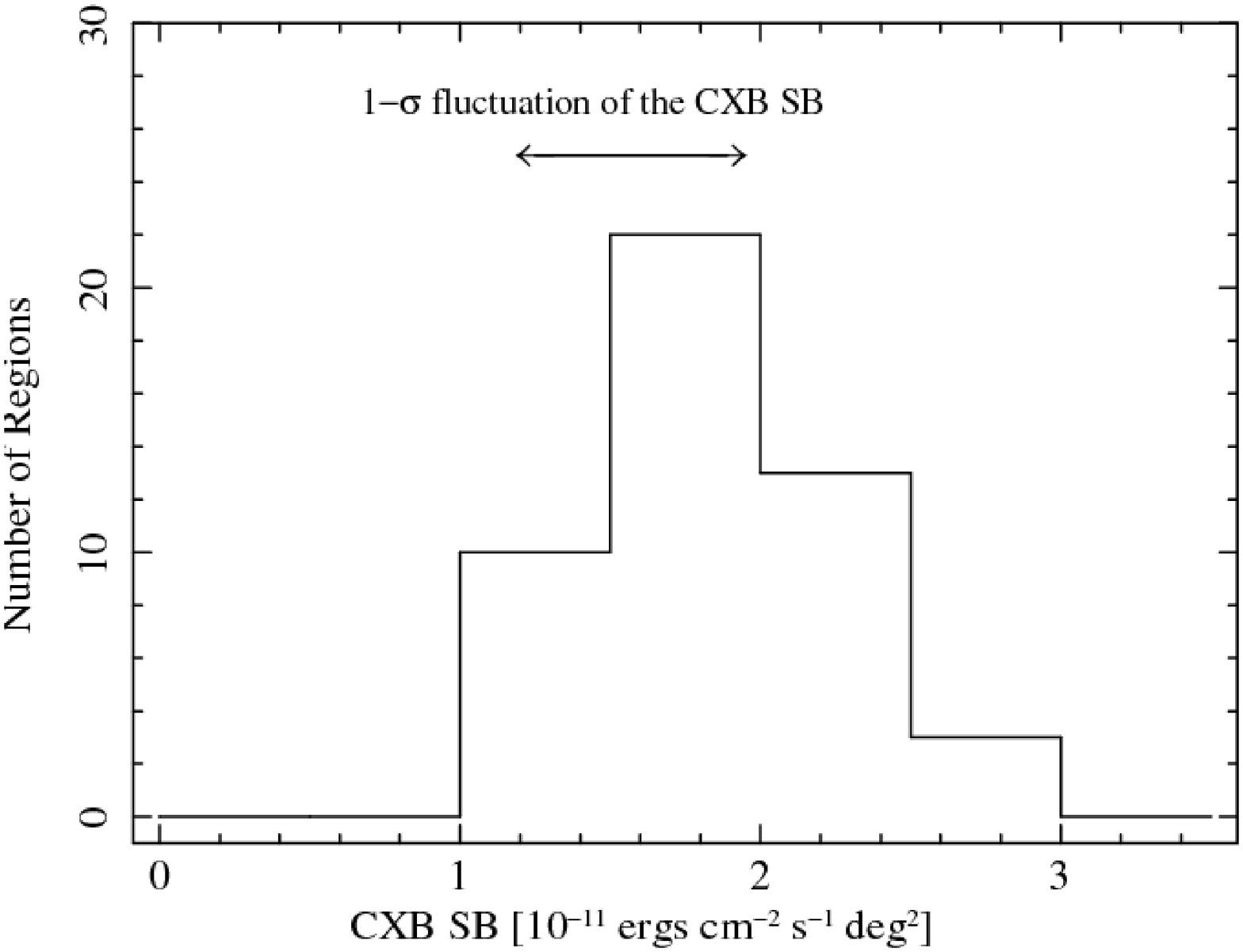}
  \end{center}
  \caption{\textit{Left}: Surface brightness (SB) distribution of the CXB component. Filled small circle, open circle, cross, and square correspond to the results at E\_rim, East, Center, and NW, respectively. Green (left) corresponds to the results at S\_rim. Black and Red triangles (right) correspond to the results in the NW\_rim,  except in the regions of 1$\arcmin$.2 and 4$\arcmin$.0 around 2XMM J045637.2+522411, respectively. The vertical error bars correspond to 1$\sigma$ errors. \textit{Right}: Histogram of the surface brightnesses. The horizontal arrow represents the 1 $\sigma$ spatial fluctuation of the CXB surface brightness estimated by \citet{Tawa08phd}.}\label{fig:flux}
\end{figure*}

\section{Spectral Analysis}
For the following analysis, the energy range of 0.4-8.0 keV was used for XIS1 (back-illuminated CCD; BI CCD) and XIS0 and XIS3 (front-illuminated CCDs; FI CCDs), since there was no signal above 8.0 keV after the background subtraction (described below).
In order to generate response matrix files (RMFs) and ancillary response files (ARFs), we employed \textit{xisrmfgen} and \textit{xissimarfgen} \citep{Ishisaki07}.
We used XSPEC version 12.6.0 \citep{Arnaud96} for all the following spectral analyses.
We excluded 2XMM J045637.2+522411 and three other point-like sources identified from the XIS image (0.5-5.0 keV).
The extracted regions are shown in figure \ref{fig:image}.

\subsection{Background Subtraction}
Since G156.2+5.7 is a large diffuse source and the FoV is filled with the SNR's emission, it is hard to obtain the background spectra from inside the FoV. 
We also have no background data from the neighborhood of the G156.2+5.7. 
For the following analysis, we used a generated non X-ray background (NXB) spectra by employing \textit{xisnxbgen} \citep{Tawa08}.
We also took into account the contributions of the local hot bubble (LHB) and the cosmic X-ray background (CXB) by assuming literature-based models.
For the LHB, we used an \textbf{APEC} model \citep{Smith01} with solar abundances \citep{Anders89} and an electron temperature of 0.1 keV \citep{Miller08}.
For the CXB we employed a broken power-law model with photon indices of $\Gamma$=2.0 ($<$0.7 keV) and $\Gamma$=1.4 ($>$0.7 keV) as shown by \citet{Miller08}.
Hereafter, we call this model the CXB model.
We also reviewed the effect of the Galactic Ridge X-ray emission (GRXE). 
The GRXE surface brightness at $l = 100\arcdeg$, $|b| <1\arcdeg$ is $\sim1.4\times10^{-12}$ erg cm$^{-2}$s$^{-1}$deg$^{-2}$ in the 0.7-10.0 keV band \citep{Sugizaki01}. 
Since G156.2+5.7 is located outside the FoV of \citet{Sugizaki01} and away from the galactic plane, the contribution of the GRXE is expected to be negligible.
For example, GRXE surface brightness  above 2.0 keV decreases about one-fortieth from $b = 0\arcdeg$ to $b \sim 9\arcdeg$ \citep{Revnivtsev06}.
Additionally, the average 0.7-10.0 keV band surface brightness of G156.2+5.7 is $\sim4.3\times10^{-9}$ erg cm$^{-2}$s$^{-1}$deg$^{-2}$ \citep{Yamauchi99} which is an order of magnitude higher than the minimum Suzaku surface brightness of $\sim3.4\times10^{-10}$ erg cm$^{-2}$s$^{-1}$deg$^{-2}$ that we see in the S\_rim observation.
In both cases, we conclude that the effect of the GRXE is negligible in G156.2+5.7.

Table \ref{tab:cxb} shows the surface brightness of the CXB and the GRXE near G156.2+5.7 and within a few degrees using observations of IRAS 05262+4432 (narrow-line Seyfert 1; \cite{Kawaguchi08}) and SGR 0501+4516 (soft gamma repeater; \cite{Enoto09}).
Since both are point-like sources, we excluded the source regions with radii 4$\arcmin$.0 and fitted the spectra obtained from the remaining regions with the CXB model.
Comparing the surface brightness of each observation, the level of the CXB and the GRXE is uniform within the range of error.
Figure \ref{fig:flux} left shows the surface brightness distribution of the CXB component inferred from the fit to each annular region in G156.2+5.7 (the details will be argued hereinafter). 
The filled circles, open circles, crosses, and squares correspond to regions from the E\_rim, East, Center, and NW, respectively. 
Green represents regions from the S\_rim.
The result suggests that the level of the CXB (and the GRXE) inside this remnant is also unchanged within the range of error.

\subsection{Model Fit for the Rim Regions' Spectra}\label{sec:rimregion}
Since the rims are mostly dominated by an interstellar medium (ISM) component \citep{Katsuda09}, we applied a single-component \textbf{VNEI} (NEI ver.2.0; \cite{Borkowski01}) model with \textbf{TBabs} (T\"{u}bingen-Boulder ISM absorption model; \cite{Wilms00}) in XSPEC for the rim spectra.
In this model, the abundances relative to the solar values \citep{Anders89} of N, O, Ne, Mg, Si, S and Fe were allowed to vary freely and that of Ni were linked to Fe. 
The other element abundances were fixed to their solar value.
Other parameters were all free; the electron temperature $kT_e$, the ionization timescale $\tau$ (a product of the electron density and the elapsed time after the shock heating), and the emission measure ($= \int n_e n_{\rm H} dl$, where $n_e$ and $n_{\rm H}$ are the number densities of hydrogen and electrons and $dl$ is the plasma depth). 
We also allowed the column density $N{\rm_H}$ to vemission measureary freely.
For the hard X-ray tail \citet{Katsuda09} applied the CXB model plus an additional power law component.
They used the same value of $N{\rm_H}$ for the CXB as for other components. 
In our analysis, it is fixed to 3.6$\times$10$^{21}$cm$^{-2}$ calculated by \textit{nh} FTOOL\footnote{http://heasarc.gsfc.nasa.gov/cgi-bin/Tools/w3nh/w3nh.pl} and separated from the $N{\rm_H}$ for G156.2+5.7.
The additional power law component was not applied for our analysis unless the CXB surface brightness exceeded the literature value.

\begin{figure}[t]
  \begin{center}
   \FigureFile(80mm,80mm){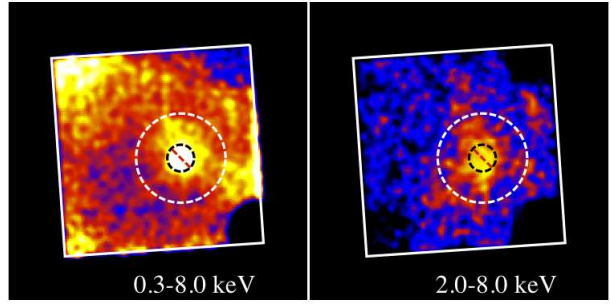}
  \end{center}
  \caption{Full-band (0.3-8.0 keV; left), and hard-band (2.0-8.0 keV; right) images of NW\_rim. The FoV of the XIS are shown with white squares. The radii of small and large circles are 1$\arcmin$.2 and 4$\arcmin$.0, respectively. 
   }\label{fig:1.2-4.0-rgb}
\end{figure}

\begin{figure}[!t]
  \begin{center}
   \FigureFile(60mm,60mm){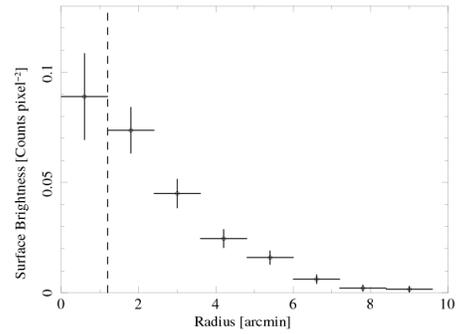}
  \end{center}
  \caption{Radial profile of NW\_rim in the 2.0-8.0 keV energy band. The center corresponds to the position of 2XMM J045637.2+522411. The dotted vertical line represents the radius of 1$\arcmin$.2.}\label{fig:radiprofile}
\end{figure}

\begin{figure}[t]
  \begin{center}
   \FigureFile(60mm,60mm){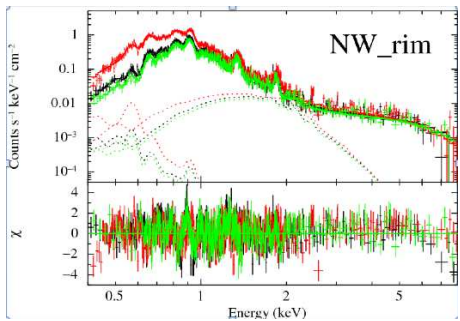}
  \end{center}
  \caption{NW\_rim spectrum fitted with the single-component VNEI model.}\label{fig:spec_inside_entire}
\end{figure}

\citet{Katsuda09} found excess emission over the CXB above 3 keV in the NW\_rim and argued that the observed excess originates from a non-thermal emission from G156.2+5.7.
In their analysis, they excluded a region of 1$\arcmin$.2 in radius where 2XMM J045637.2+522411 is located.
Although \citet{Yamauchi10} estimated the extension of the X-ray emission from 2XMM J045637.2+522411 to be at least $\sim$50$\arcsec$ (the surface brightness within it is $\sim$2.64$\times$10$^{-10}$ erg cm$^{-2}$s$^{-1}$deg$^{-2}$ in the 2.0-10.0 keV band) based on the XMM-Newton data, we noticed that its extent might be larger.
Figure \ref{fig:1.2-4.0-rgb} left and right panels show a full-band (0.3-8.0 keV) and a hard-band (2.0-8.0 kev) image, respectively.
In figure \ref{fig:1.2-4.0-rgb}, we show the small circular region of 1$\arcmin$.2 in radius as used by \citet{Katsuda09}.  We find that  the X-ray emission from the cluster extends beyond this circle even in the hard band.
Figure \ref{fig:radiprofile} shows a radial profile from the center of 2XMM J045637.2+522411 in the 2.0-8.0 keV band. 
The dotted vertical line corresponds to the radius of 1$\arcmin$.2.
Figure \ref{fig:radiprofile} shows that the surface brightness decreases to half at the radius of $\sim$3$\arcmin$.
To rule out the possible contamination from 2XMM J045637.2+522411, we therefore excluded a circular region with radius 4$\arcmin$.0 as shown with the white large circle in figure \ref{fig:1.2-4.0-rgb}.
The fitting result after excluding this larger region is shown in figure \ref{fig:spec_inside_entire}.
The best-fit parameters are shown in table \ref{tab:spec_inside_entire}.


The surface brightness of the hard component in the NW\_rim is calculated to be 1.88$\times$10$^{-11}$ erg cm$^{-2}$s$^{-1}$deg$^{-2}$ (2.0-10.0 keV; table \ref{tab:spec_inside_entire}).
On the other hand, the CXB surface brightness is estimated to be 1.57$\pm$0.32$\times$10$^{-11}$ erg cm$^{-2}$s$^{-1}$deg$^{-2}$ based on blank sky observations with the XIS \citep{Tawa08phd}.
In their analysis, a spatial fluctuation 14\% (1$\sigma$) is considered for the XIS FoV \citep{Tawa08}.
Accordingly, the surface brightness of the hard component in the NW\_rim (and also E\_rim and S\_rim) is within the uncertainty of the CXB surface brightness.
We conclude that no additional power law component is required for our fit.

\begin{table*}[hbt]
  \caption{Spectral parameters}\label{tab:spec_inside_entire}
  \begin{center}
    \begin{tabular}{llccc}
       \hline 
      \hline
Component & Parameter & \multicolumn{3}{c}{Value} \\
      \hline
 & &  NW\_rim & E\_rim & S\_rim \\
      CXB model\\
      \ \ \textit{absorption} & N$\rm _H$ [10$^{21}$ cm$^{-2}$]\dotfill & \multicolumn{3}{c}{3.6 (fixed)} \\
      \ \ \textit{Broken Power-law}  & $\Gamma_1$\dotfill & \multicolumn{3}{c}{2.0 (fixed)} \\
       & E$\rm_{break}$ [keV]\dotfill & \multicolumn{3}{c}{0.7 (fixed)} \\
       & $\Gamma_2$\dotfill & \multicolumn{3}{c}{1.4 (fixed)} \\
       & surface brightness\footnotemark[$*$]\dotfill & 1.88$\times$10$^{-11}$ & 1.61$\times$10$^{-11}$ & 1.56$\times$10$^{-11}$\\
      LHB model\\
      \ \ \textit{APEC} & $kT_e$ [keV]\dotfill & \multicolumn{3}{c}{0.1 (fixed)} \\
       & Abundances\dotfill & \multicolumn{3}{c}{1 (fixed)} \\
       & surface brightness\footnotemark[$*$]\dotfill & \multicolumn{3}{c}{3.3$\times$10$^{-16}$ (fixed)} \\
      VNEI model\\
      \ \ \textit{absorption} & N$\rm _H$ [10$^{21}$ cm$^{-2}$]\dotfill &   2.3$\pm$0.1 & 2.0$\pm$0.1 & 3.1$\pm$0.1  \\
       &  $kT_e$ [keV]\dotfill & 0.45$\pm$0.01 &  0.49$\pm$0.01 & 0.41$\pm$0.01 \\
      & log($\tau$ [s cm$^{-3}$])\dotfill &   10.64$\pm$0.02 & 10.54$^{+0.01}_{-0.02}$ & 10.80$^{+0.01}_{-0.03}$\\
      & N\dotfill &    0.12$\pm$0.01 & 0.15$\pm$0.01 & 0.50$\pm$0.01\\
      & O\dotfill & 0.24$\pm$0.01 & 0.22$\pm$0.01 & 0.25$\pm$0.01\\
      & Ne\dotfill        & 0.42$\pm$0.01 & 0.44$\pm$0.01 & 0.57$\pm$0.01 \\
      & Mg\dotfill        & 0.37$\pm$0.02 & 0.34$\pm$0.01 & 0.47$\pm$0.01\\
      & Si (=S)\dotfill       & 0.71$\pm$0.08 & 0.32$\pm$0.01 & 0.48$\pm$0.14\\
      & Fe (=Ni)\dotfill   & 0.36$\pm$0.01 & 0.37$\pm$0.01 & 0.40$\pm$0.01\\
 \\
&  $\chi ^2$/dof\dotfill    & 1395/1245 & 1441/1190 & 1183/1031 \\
      \hline
\multicolumn{3}{l}{\footnotemark[$*$]{ 2.0-10.0 keV surface brightness in units of erg cm$^{-2}$s$^{-1}$deg$^{-2}$}} \\
\multicolumn{3}{l}{Other elements are fixed to solar values.}\\
\multicolumn{3}{l}{The errors are in the range $\Delta\chi^{2}<2.7$ on one parameter.}\\
    \end{tabular}
 \end{center}
\end{table*}

\subsection{Model Fit for the Inner Regions' Spectra}
We also applied the single-component VNEI model for each inner region. 
However, this model was inadequate to fit the data due to its simplicity. 
As previously shown by \citet{Katsuda09}, the thermal emission from the interior of G156.2+5.7 consists of two components with different origins; a high-temperature ejecta component and a low-temperature ISM component.
Thus we applied a two-component VNEI model  for the spectra obtained from Center, East and NW. 
In this model, we fixed the ISM abundances to the average values based on the fitting results of rim observations (see table \ref{tab:spec_inside_entire}).
The average abundances of the ISM component are set as follows: N=0.23, O=0.24, Ne=0.45, Mg=0.39, Si(=S)=0.50, Fe(=Ni)=0.38. 
We fixed the other abundances to the solar values. 
We also set $kT_e$ to be 0.45 keV which is an average electron temperature for the rim regions of G156.2+5.7.
The values of $\tau$ and emission measure were allowed to vary.
Meanwhile, in the ejecta component, the abundances of O, Ne, Mg, Si, S and Fe were free while we set the abundances of C and N equal to O, Ni equal to Fe. 
All other abundances were fixed to their solar values.
The other free parameters were $kT_e$, $\tau$, emission measure, and the column density $N{\rm_H}$.

We note that the temperature of the low-kT component may be higher than that of the rim because the line-of-sight ISM component includes ambient medium  shocked earlier in the evolution of the remnant when the shock velocity was higher. 
In general, higher ISM temperatures will increase the line emission from species such as Si and S, which may affect the ejecta abundance measurement.
Since it is impossible to measure the real average temperature of the ISM component, we examined how the metal abundances change by increasing the ISM temperature from 0.45 keV to 2.0 keV. 
As a result, we concluded that the ISM temperature hardly affect the ejecta abundances within the range of the error.
We thus fixed the ISM temperature to be 0.45 keV for the inner regions' spectra.

\section{Discussion}
\subsection{Ejecta Structure}\label{sec:ejecta}
\begin{figure*}
  \begin{center}
    \FigureFile(45mm,45mm){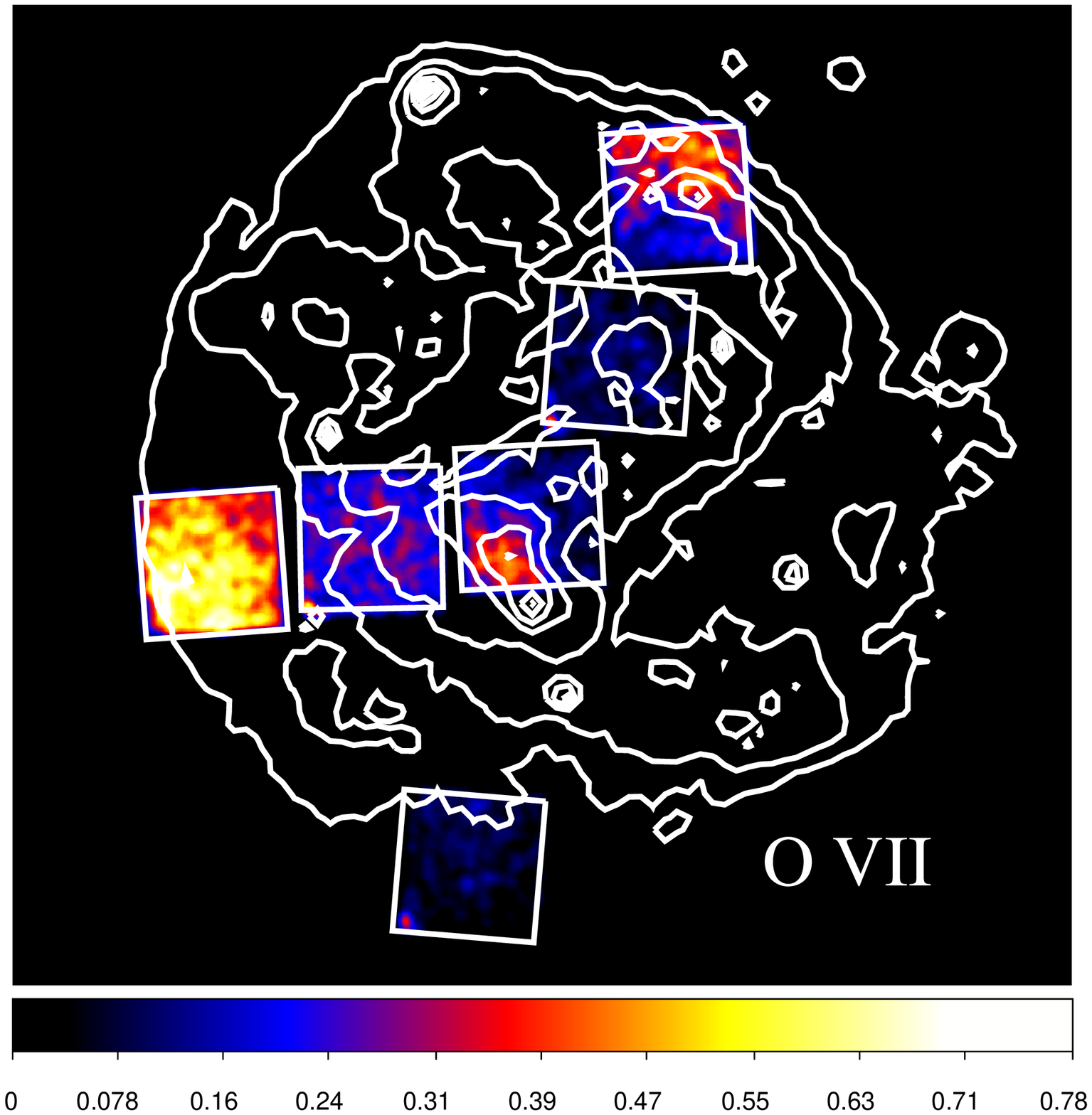}
    \FigureFile(45mm,45mm){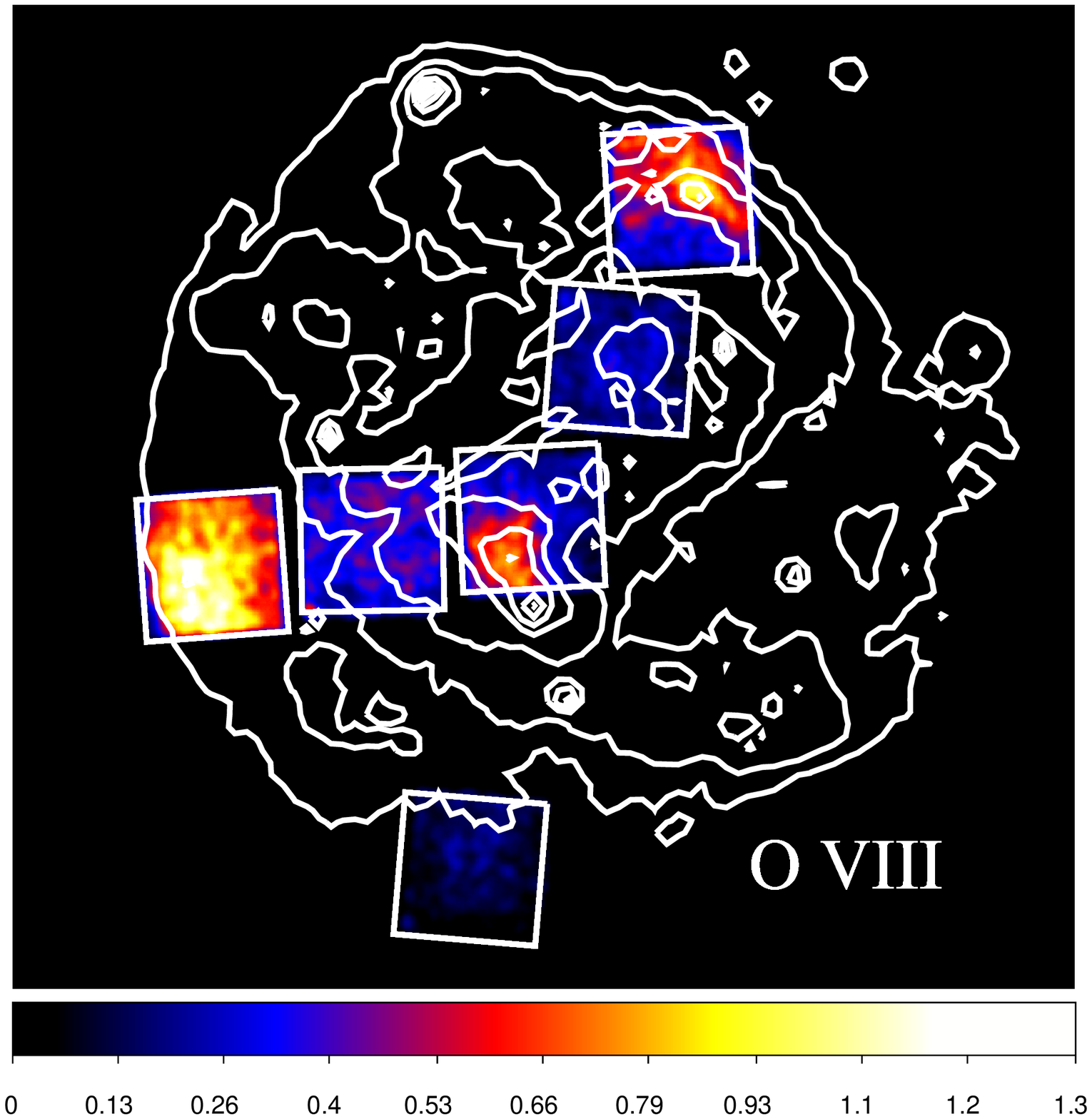}
    \FigureFile(45mm,45mm){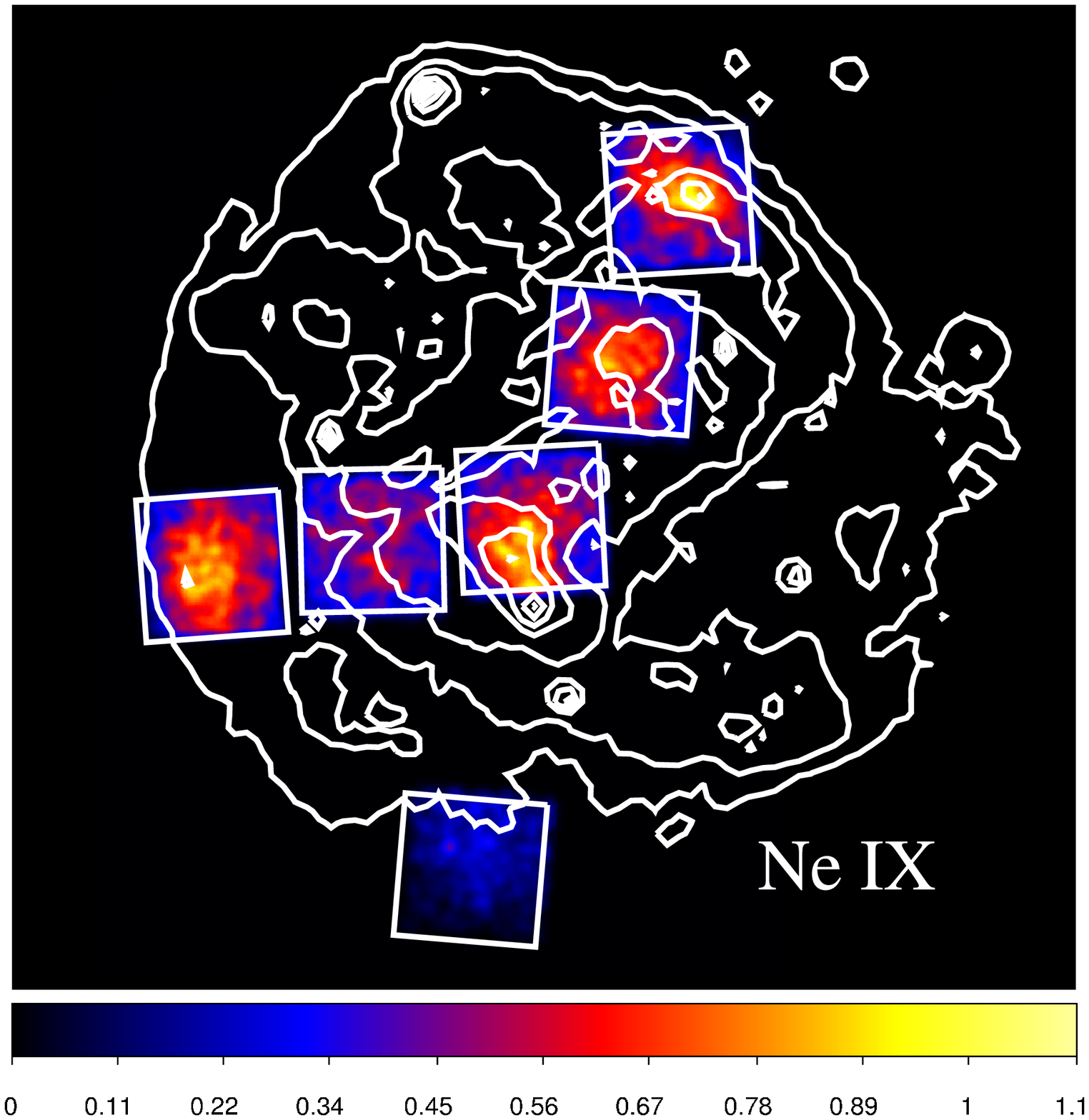}
    \FigureFile(45mm,45mm){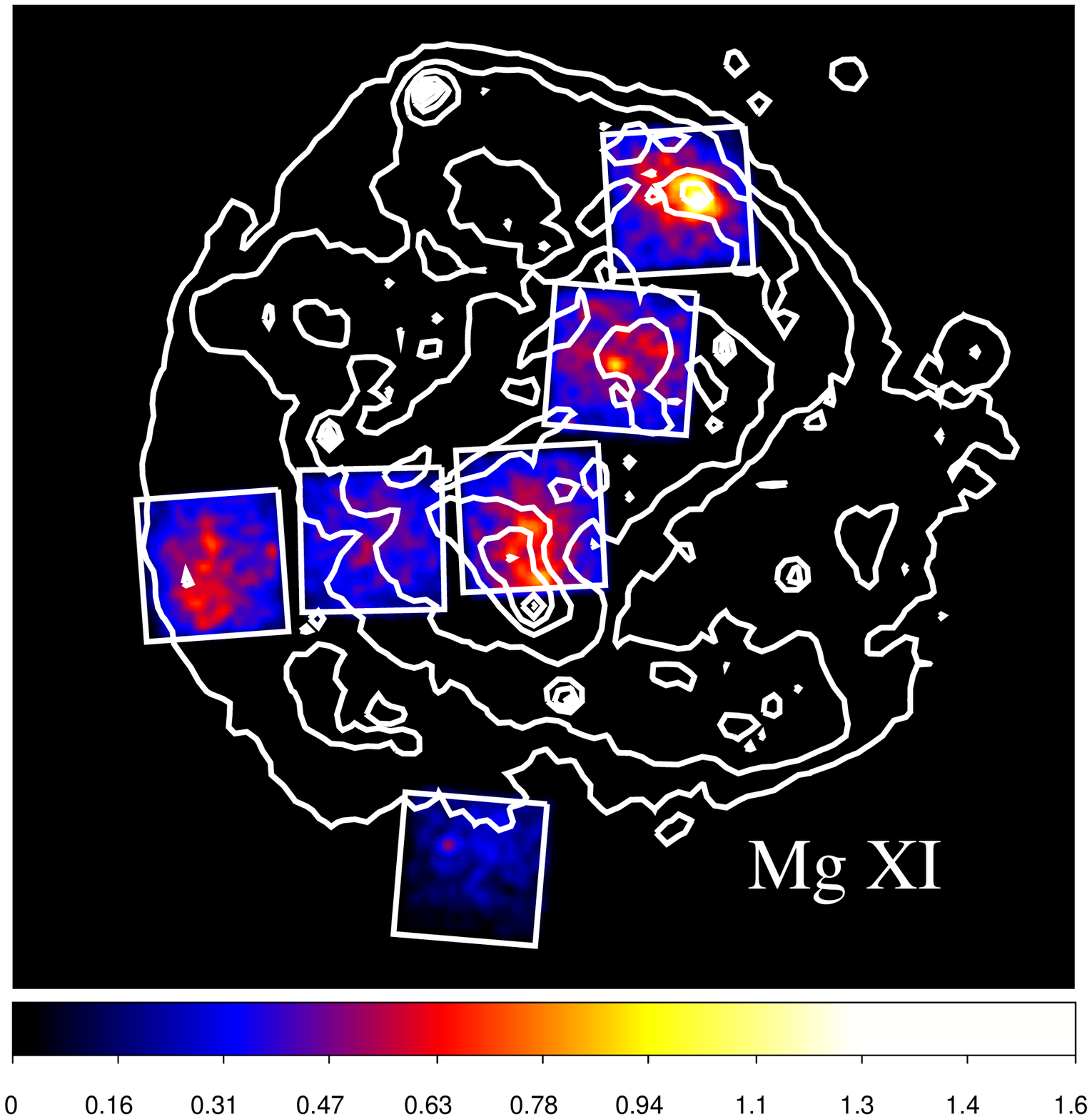}
    \FigureFile(45mm,45mm){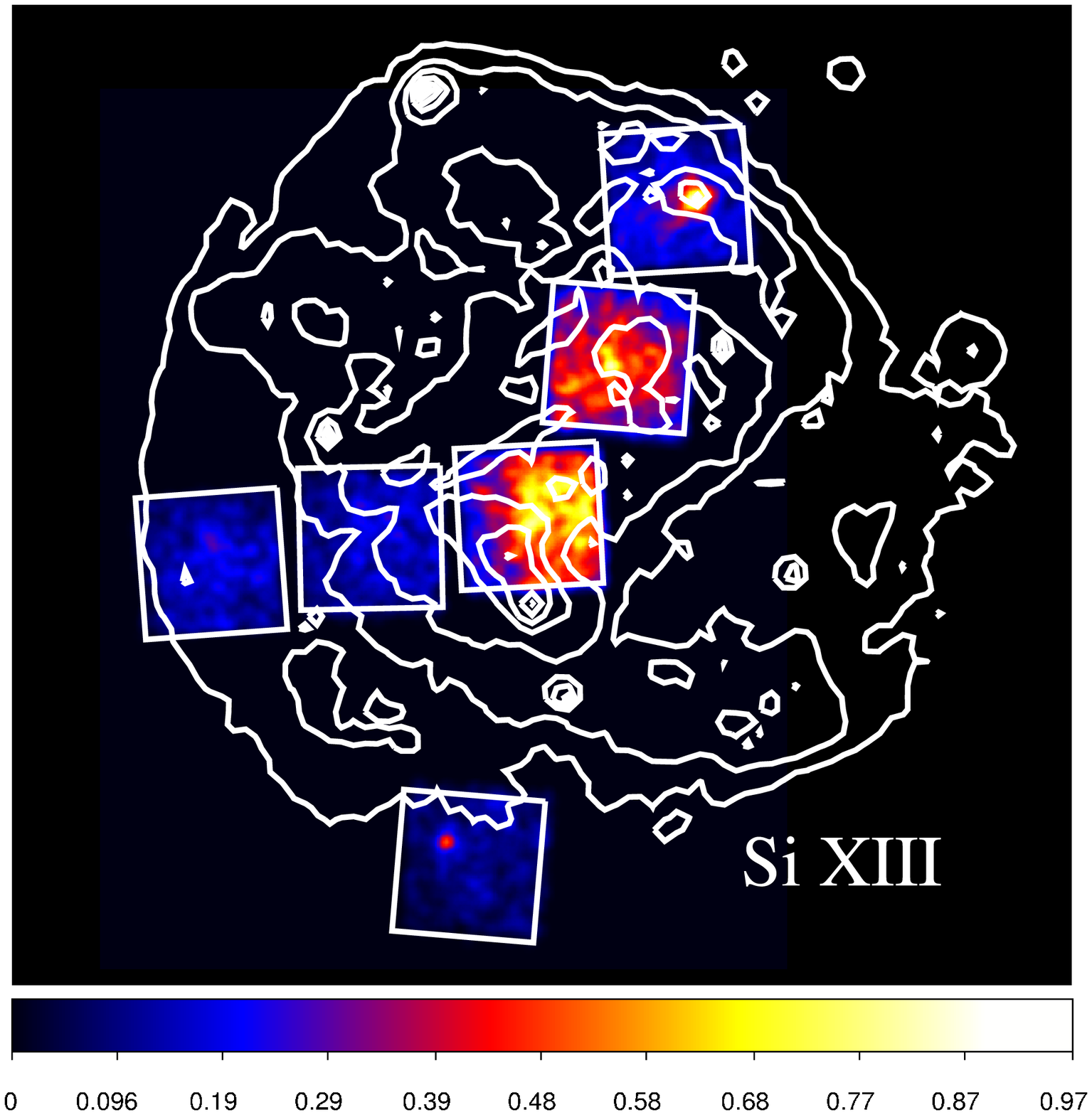}
    \FigureFile(45mm,45mm){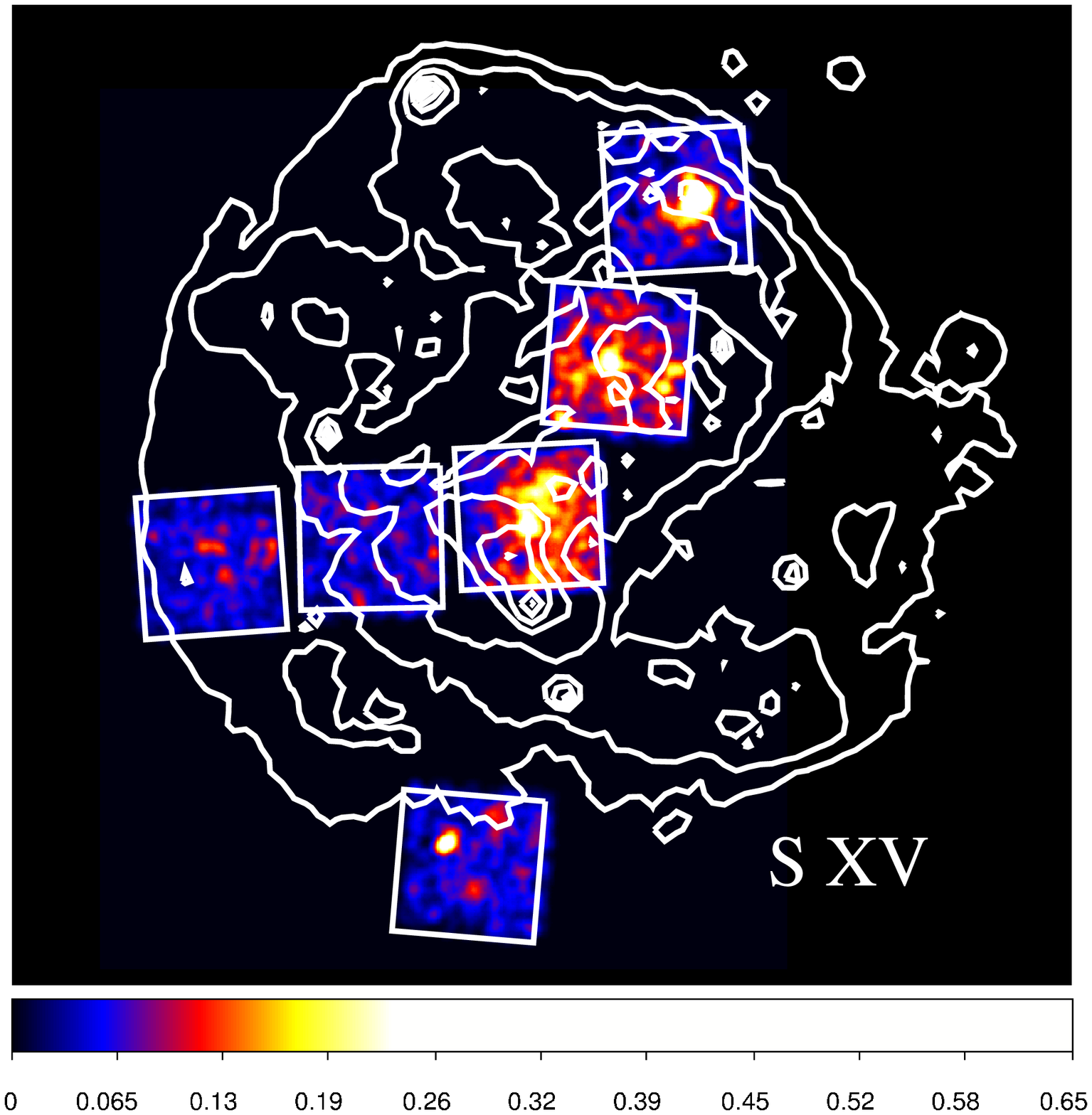}
  \end{center}
  \caption{Band images of several elements. The white contours were derived from a soft X-ray image taken by ROSAT PSPC. }\label{fig:band}
\end{figure*}
Figure \ref{fig:band} shows band images of O \emissiontype{VII}, O \emissiontype{VIII}, Ne \emissiontype{IX}, Mg \emissiontype{XI}, Si \emissiontype{XIII} and S \emissiontype{XV} using all the Suzaku data.
In figure \ref{fig:band}, both the Si and the S distributions appear to be centrally concentrated, while those of O \emissiontype{VII} and O \emissiontype{VIII} show a clear limb brightening, which suggests that the observed heavy elements originated from the ejecta.
Furthermore, figure \ref{fig:band} suggests that the distribution centers of Si and S shift toward the northwest.

\begin{figure}[!htb]
  \begin{center}
   \FigureFile(60mm,60mm){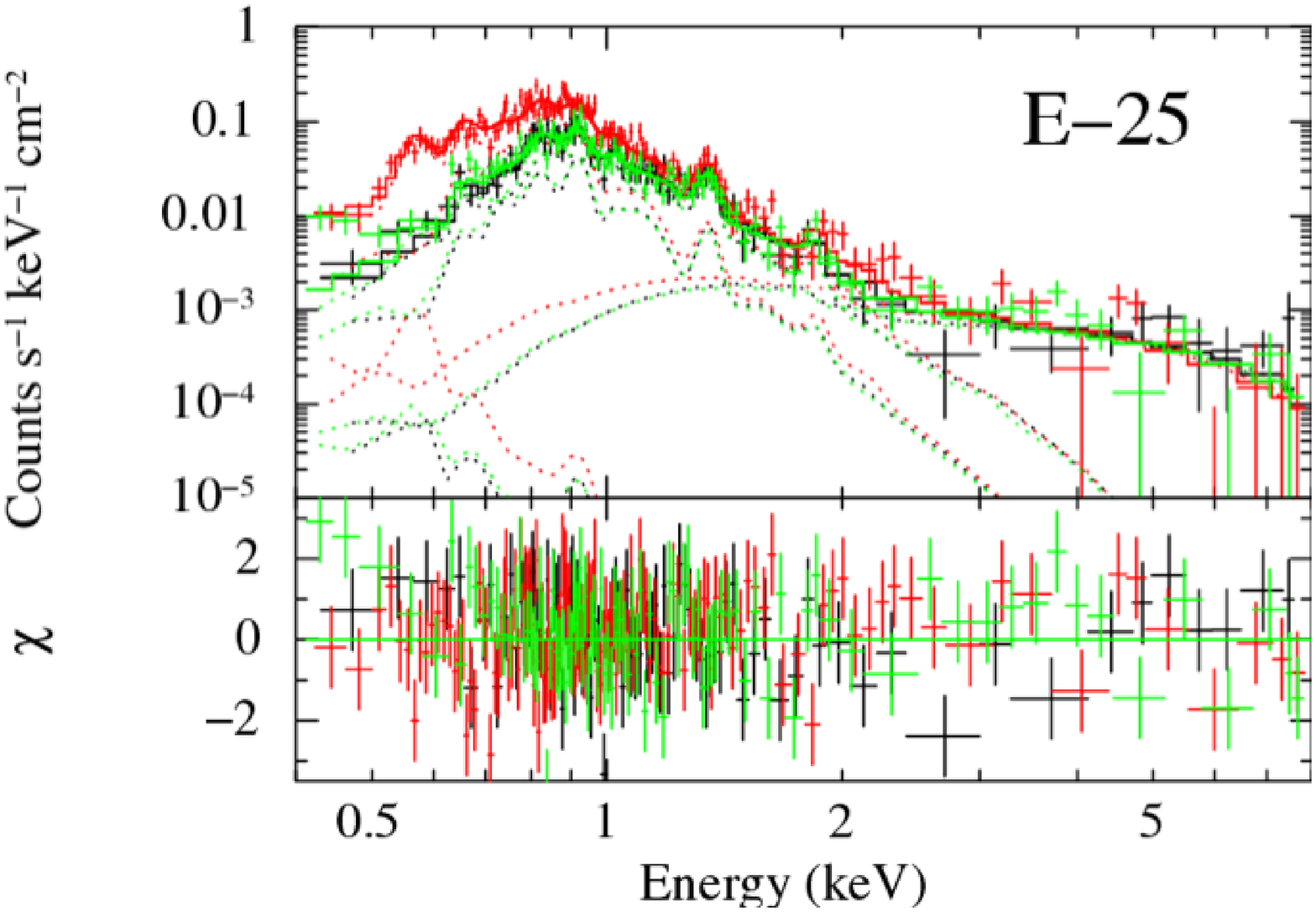}
    \FigureFile(60mm,60mm){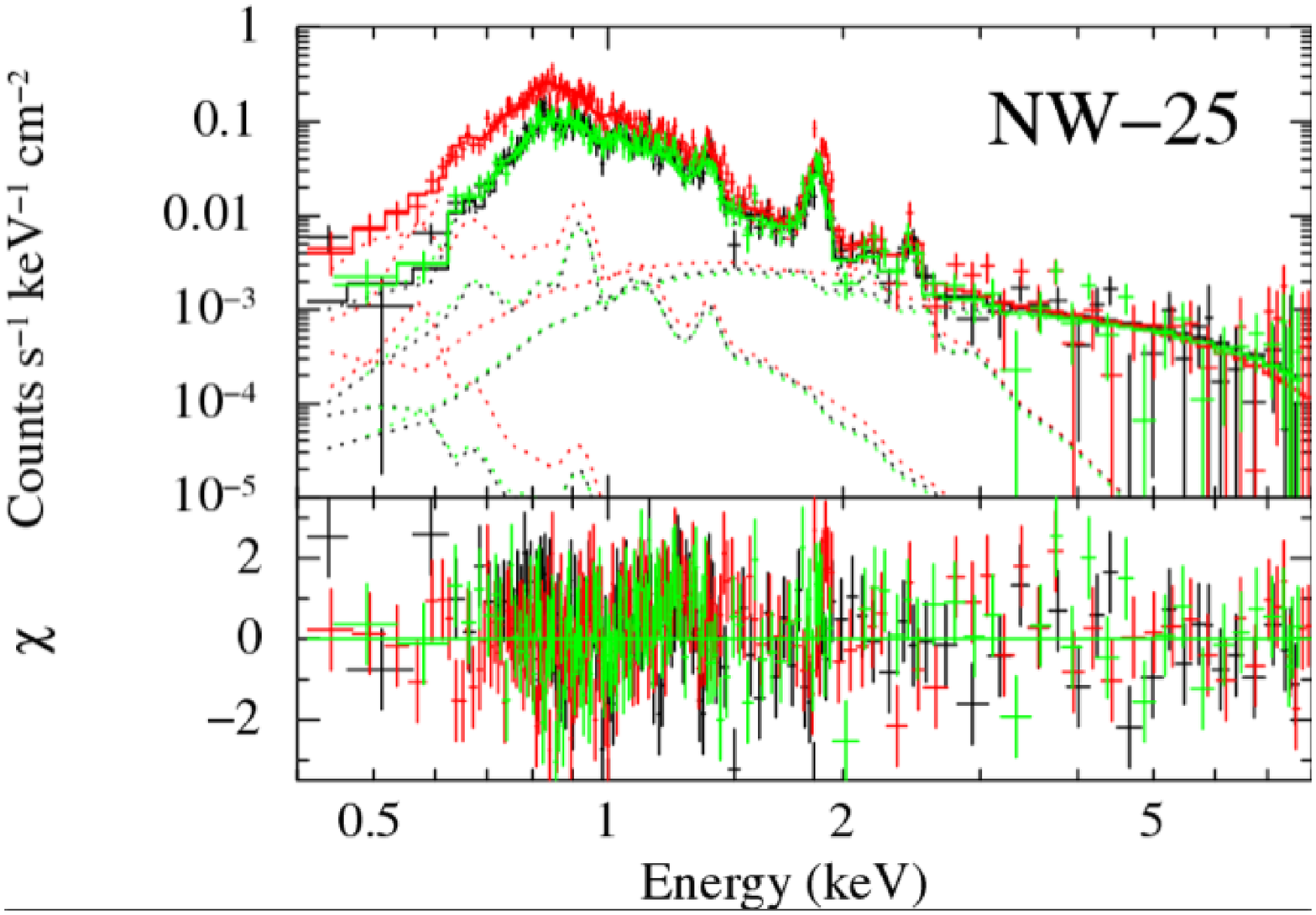}
  \end{center}
  \caption{Example spectra of two annular regions in G156.2+5.7. Both were fitted with the two-component VNEI model.}\label{fig:spec_inside}
\end{figure}

To examine the radial distribution of the plasma, we divided the data into annular regions, as shown in figure \ref{fig:image} right. 
As the annular center we used 04$^{\mathrm h}$58$^{\mathrm m}$54$^{\mathrm s}$.9, 51\arcdeg44\arcmin36\arcsec.1 (J2000), the optical axis during the observation of Center.
We used an annular width of 2$\arcmin$,  consistent with  the angular resolution of the XIS.
In order to estimate the boundary of the ejecta-dominated region, we applied both the single-component VNEI model and the two-component VNEI model to the spectra obtained from the E\_rim and NW\_rim.
We determined which model is preferred by using the F-test with a significance level of 99\%.
We found that the second component is required in the NW\_rim only within roughly $\sim$85\% of the shock radius ($\sim$52$\arcmin$).
Figure \ref{fig:spec_inside} shows example spectra obtained from the annular regions 25\arcmin \ away from the annular center toward the east (E-25; left) and the northwest (NW-25; right).
The best-fit parameters are shown in table \ref{tab:spec_inside}.
While both spectra require the two-component VNEI model, the abundances of the ejecta component are significantly different particularly for Si and S.
We also note that there is no positive evidence that G156.2+5.7 is an O-rich SNR as claimed by \citet{Hudaverdi10}.
Their argument is that the south and east side of their FoV are soft X-ray dominated (0.4-1.0 keV), suggesting the abundant presence of O.
However, we consider that such excesses are mainly due to the contribution of the low-temperature ISM component.

\begin{table*}
  \caption{Spectral parameters (see figure \ref{fig:spec_inside})}\label{tab:spec_inside}
  \begin{center}
    \begin{tabular}{llcc}
       \hline 
      \hline
Component & Parameter & \multicolumn{2}{c}{Value} \\
      \hline
 & & Region E-25 & Region NW-25 \\
      CXB model\\
      \ \ \textit{absorption} & N$\rm _H$ [10$^{21}$ cm$^{-2}$]\dotfill & \multicolumn{2}{c}{3.6 (fixed)} \\
      \ \ \textit{Broken Power-law}  & $\Gamma_1$\dotfill & \multicolumn{2}{c}{2.0 (fixed)} \\
       & E$\rm_{break}$ [keV]\dotfill & \multicolumn{2}{c}{0.7 (fixed)} \\
       & $\Gamma_2$\dotfill & \multicolumn{2}{c}{1.4 (fixed)} \\
       & surface brightness\footnotemark[$*$]\dotfill & 1.48$\times$10$^{-11}$ & 1.74$\times$10$^{-11}$ \\
      LHB model\\
      \ \ \textit{APEC} & $kT_e$ [keV]\dotfill & \multicolumn{2}{c}{0.1 (fixed)} \\
       & Abundances\dotfill & \multicolumn{2}{c}{1 (fixed)} \\
       & surface brightness\footnotemark[$*$]\dotfill & \multicolumn{2}{c}{3.3$\times$10$^{-16}$ (fixed)} \\
      VNEI model\\
      \ \ \textit{absorption} & N$\rm _H$ [10$^{21}$ cm$^{-2}$]\dotfill & 1.8$\pm$0.1 & 3.4$\pm$0.1 \\
      \ \ \textit{Low-$kT_e$ component} & $kT_e$ [keV]\dotfill & \multicolumn{2}{c}{0.45 (fixed)} \\
      & log($\tau$ [s cm$^{-3}$])\dotfill & 11.38$^{+0.08}_{-0.07}$ &  10.00$^{+0.09}_{-0.08}$ \\
      & N\dotfill &  \multicolumn{2}{c}{0.23 (fixed)} \\
      & O\dotfill &  \multicolumn{2}{c}{0.24 (fixed)}  \\
      & Ne\dotfill       &  \multicolumn{2}{c}{0.45 (fixed)} \\
      & Mg\dotfill       &  \multicolumn{2}{c}{0.39 (fixed)} \\
      & Si\dotfill       &  \multicolumn{2}{c}{0.50 (fixed)} \\
      & S\dotfill        &  \multicolumn{2}{c}{0.50 (fixed)} \\
      & Fe (=Ni)\dotfill &  \multicolumn{2}{c}{0.38 (fixed)} \\
      \ \ \textit{High-$kT_e$ component} & $kT_e$ [keV]\dotfill &  0.59$^{+0.02}_{-0.01}$  & 0.58$\pm$0.01 \\
      &  log($\tau$ [s cm$^{-3}$])\dotfill & 11.29$\pm$0.04 & 11.14$^{+0.07}_{-0.04}$ \\
      & O (=C=N)\dotfill  &  1.18$^{+0.22}_{-0.21}$ & 0.61$\pm$0.09 \\
      &  Ne\dotfill       &  0.51$\pm$0.11 & 0.56$\pm$0.08 \\
      &  Mg\dotfill       &  0.83$^{+0.13}_{-0.12}$ & 0.63$\pm$0.08 \\
      &  Si\dotfill       &  0.48$\pm$0.22 & 2.25$\pm$0.20 \\
      &  S\dotfill        &  $<$1.5 & 3.40$\pm$0.96 \\
      &  Fe (=Ni)\dotfill &  1.05$\pm$0.07 & 1.45$\pm$0.06 \\
\\
&  $\chi ^2$/dof\dotfill  & 379/326  & 533/453 \\
      \hline
\multicolumn{2}{l}{\footnotemark[$*$]{ 2.0-10.0 keV surface brightness in units of erg cm$^{-2}$s$^{-1}$deg$^{-2}$}} \\
\multicolumn{2}{l}{Other elements are fixed to solar values.}\\
\multicolumn{2}{l}{The errors are in the range $\Delta\chi^{2}<2.7$ on one parameter.}\\
    \end{tabular}
 \end{center}
\end{table*}

\begin{figure*}
  \begin{center}
    \FigureFile(80mm,80mm){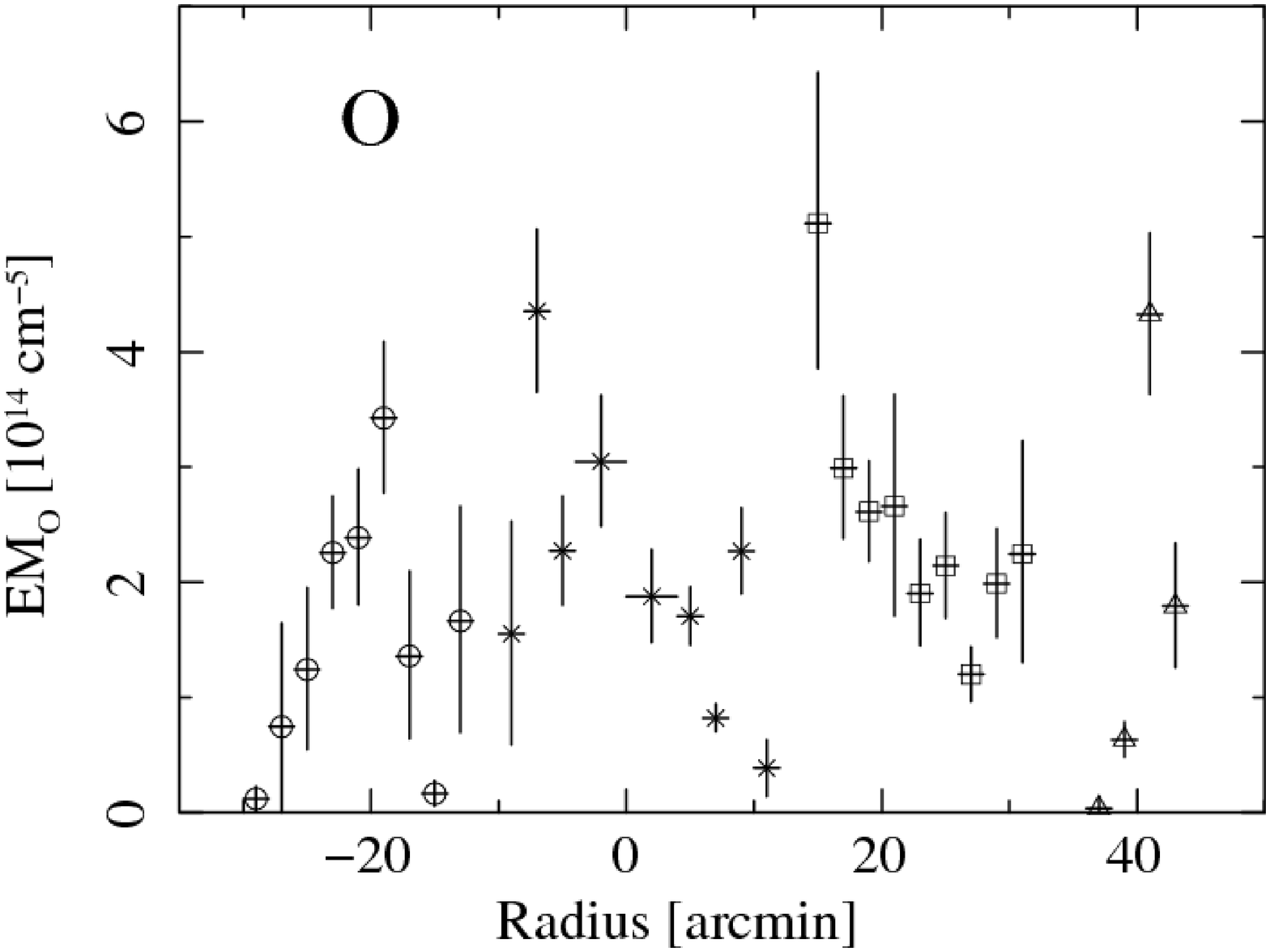}
    \FigureFile(80mm,80mm){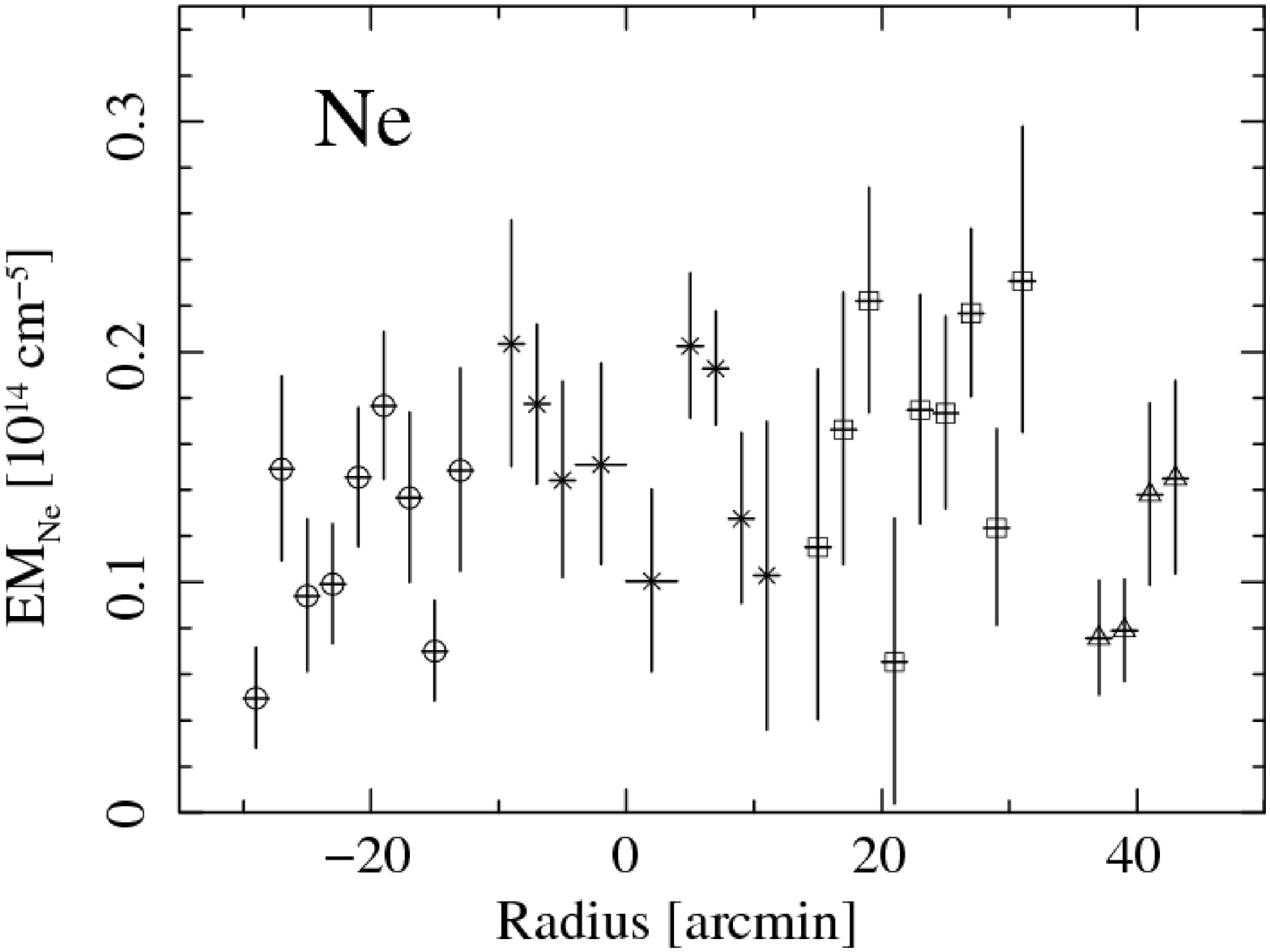}
    \FigureFile(80mm,80mm){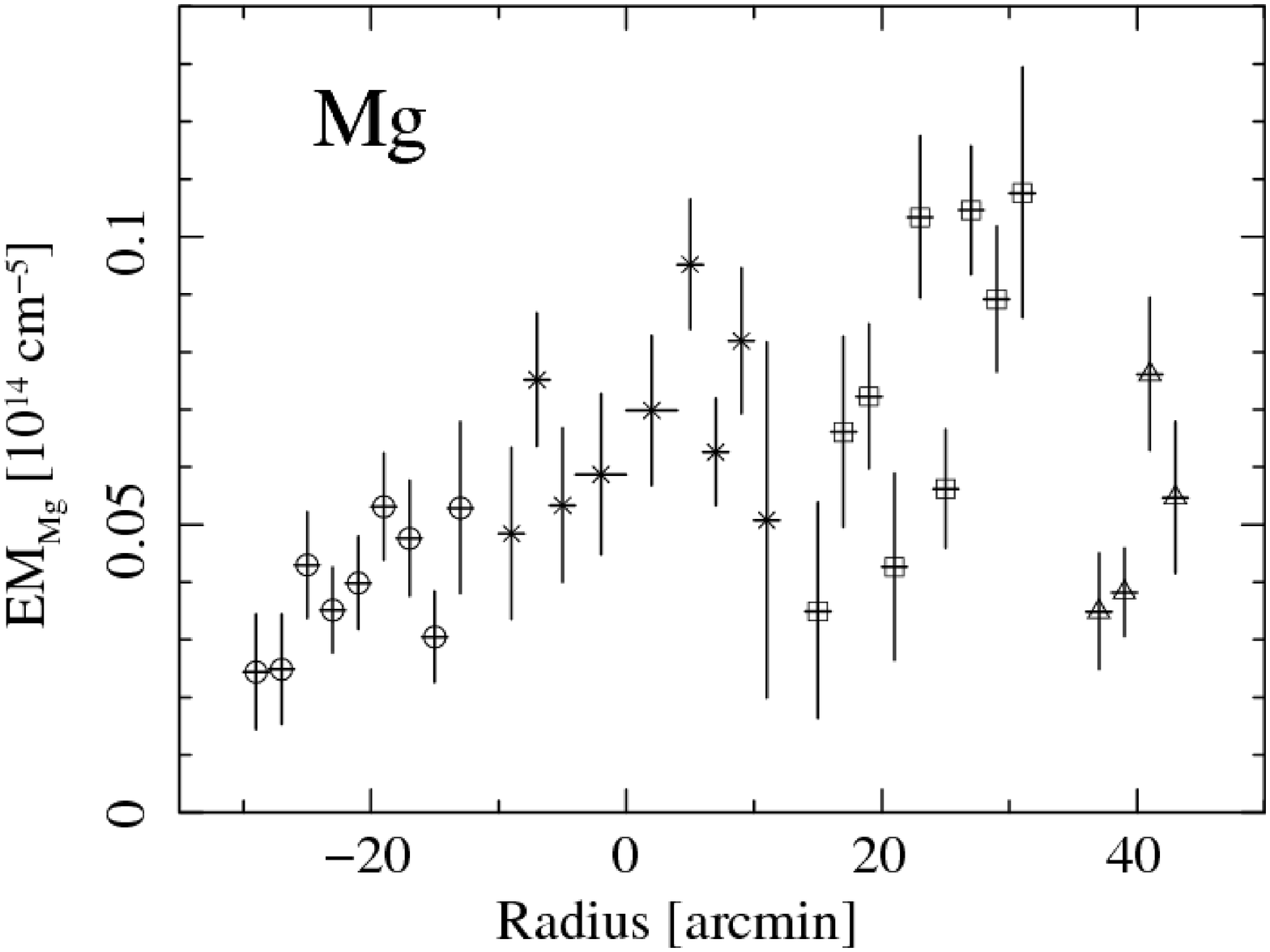}
    \FigureFile(80mm,80mm){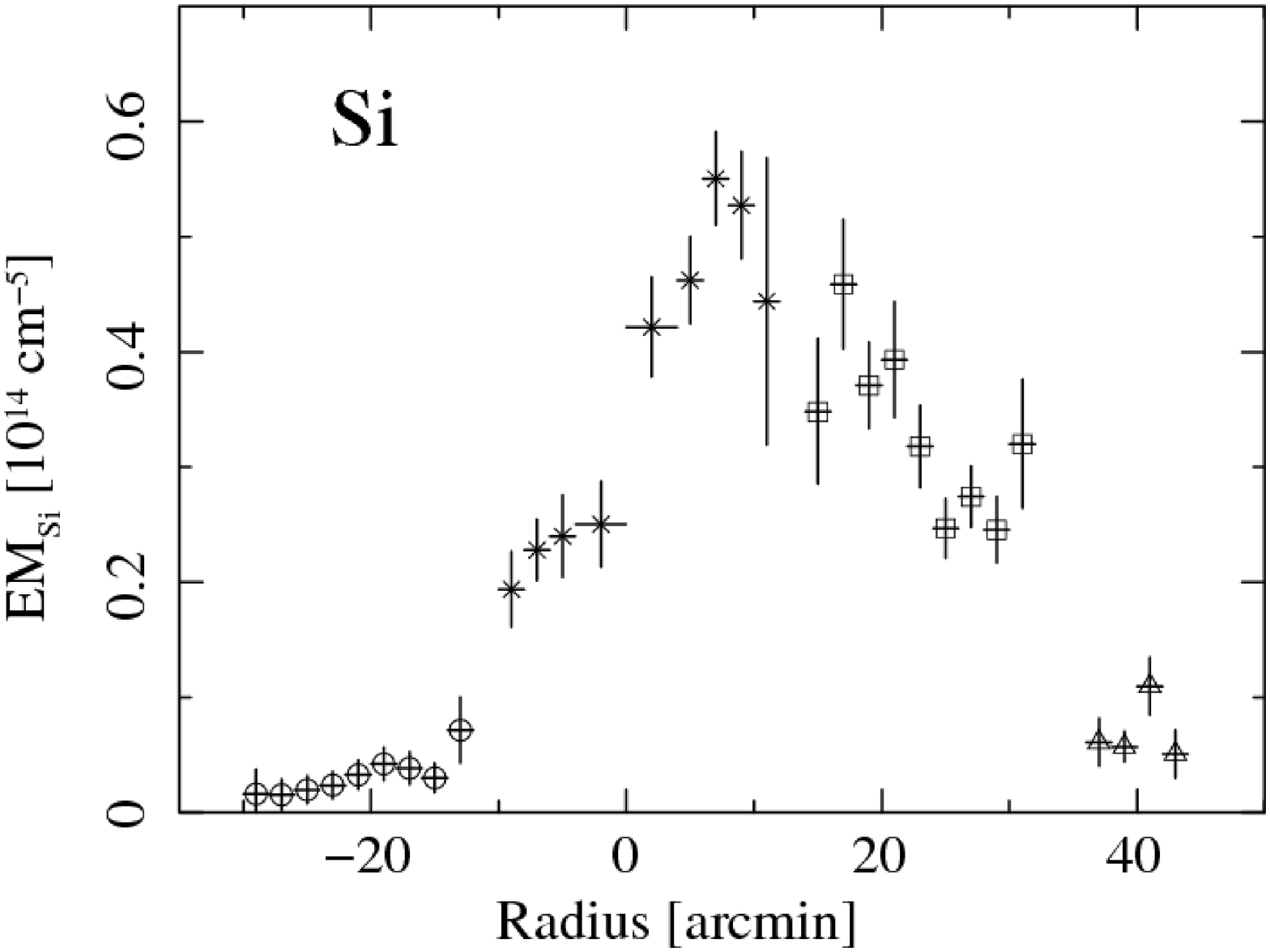}
    \FigureFile(80mm,80mm){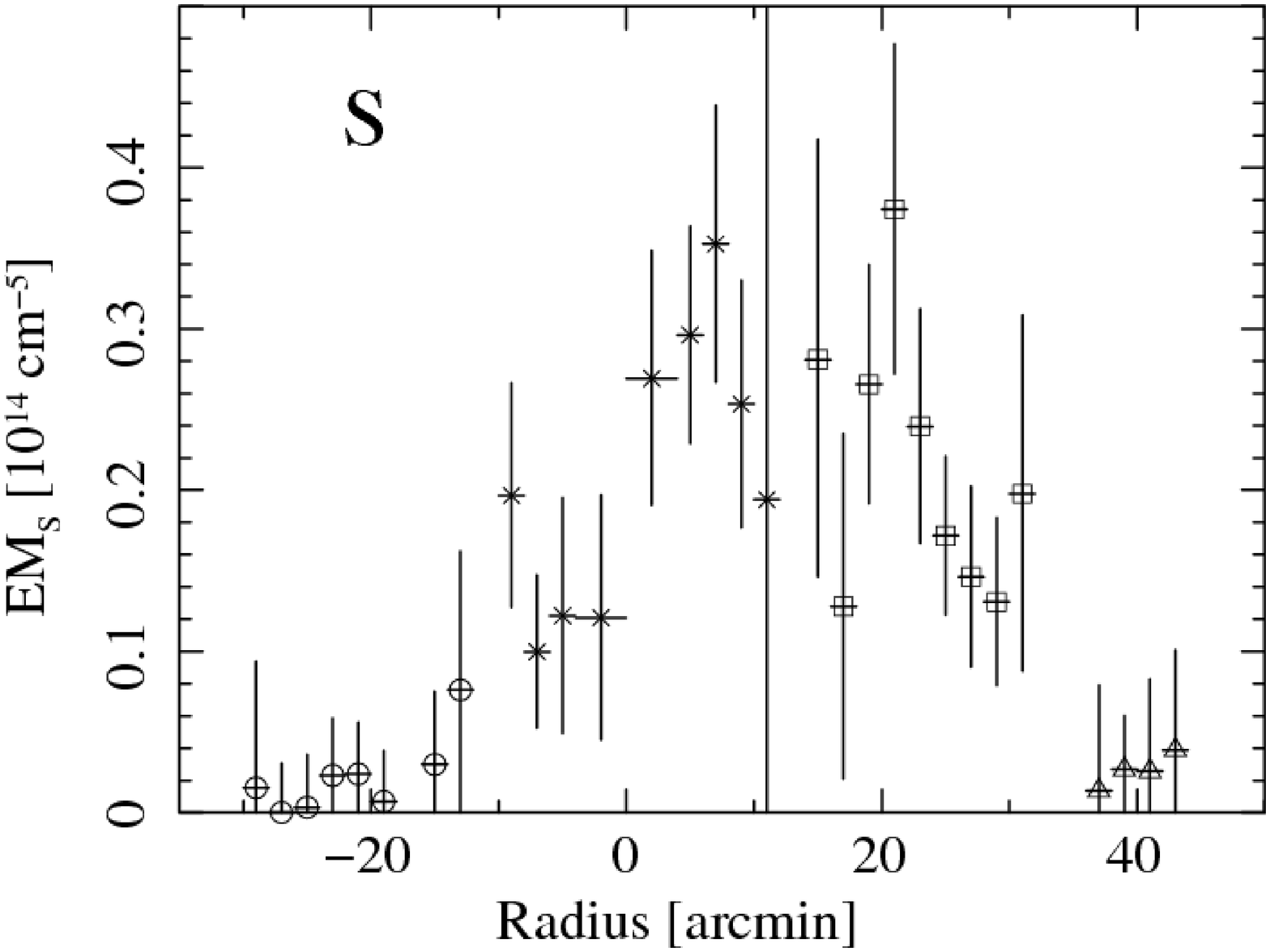}
    \FigureFile(80mm,80mm){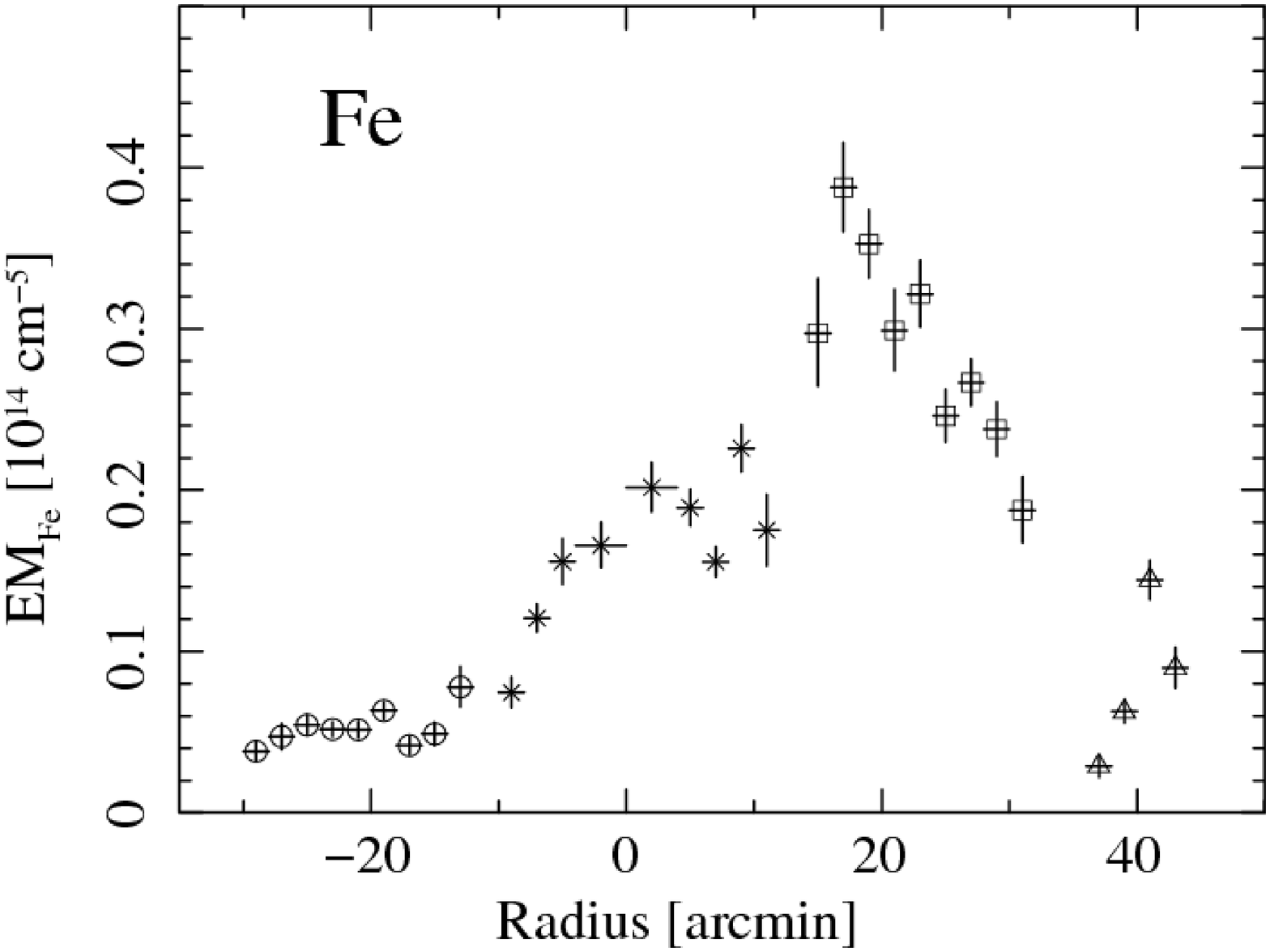}
  \end{center}
  \caption{Distributions of emission measure for various metals (O, Ne, Mg, Si, S, and Fe) in the ejecta. Open circle, cross, square, and triangle correspond to the spectral extracted regions, East, Center, NW, and NW\_rim, respectively. The horizontal axis shows a distance from the center of Center. The positive and the negative values represent directions toward the northwest and the east, respectively.}\label{fig:emission measure}
\end{figure*}

Figure \ref{fig:emission measure} shows the radial profiles of the emission measure for various metals (O, Ne, Mg, Si, S, and Fe) in the ejecta.
We found prominent central concentrations for Si, S, and Fe, while the lighter elements (O, Ne, and Mg) are distributed more uniformly.
The emission measure of Si peaks $\sim$10$\arcmin$ from the annular center and the overall distribution is quite similar to that of S.
The co-existence between Si and S is consistent with a theoretical picture that they are mainly produced in the static O-burning layer during the stellar evolution and the incomplete Si-burning layer during the SN explosion.
Similar shifts are observed in other evolved SNRs, for example, the Cygnus Loop \citep{Uchida09}.
If such inhomogeneities are real, one of the possible causes is an asymmetric explosion of a progenitor star.
While many SN models have been proposed in recent decades, a model invoking delayed explosion caused by neutrino heating is worth considering in the context of these data.
For example, \citet{Marek09} simulated an SN explosion of 15\MO \ caused by the SASI-aided core-collapse (SASI; standing accretion shock instability).
They expect that the heavy elements, such as Si and Fe, are ejected in one direction, while the lighter elements, such as O, Ne, and Mg, are ejected in all directions.
However, since our FoV is limited, we cannot determine the precise distribution center of each ejected element.
Also, the annular center does not necessarily correspond to the explosion center.
It should be carefully studied whether the origin of such inhomogeneous distribution is actually the result of an asymmetric SN explosion or not.

\subsection{Origin of hard X-ray emission}\label{sec:hardxray}
As shown in section \ref{sec:rimregion}, we found no robust evidence for non-thermal emission in the NW\_rim when we excluded the region contaminated by the cluster 2XMM J045637.2+522411.
The consistency of the measured CXB surface brightness from the NW\_rim with other observations' measurements (see table \ref{tab:cxb}) as well as with values found in the literature strengthens the conclusion that there is no non-thermal emission from the NW\_rim.

We next investigate spatial fluctuations of the hard X-ray surface brightness by using all the spectra obtained from the annular regions described above (see figure \ref{fig:flux}).
From figure \ref{fig:flux} left, there is no correlation between the radio synchrotron and the hard X-ray emission contrary to the prediction by \citet{Katsuda09}.
The hard X-ray distribution does not have a bilaterally symmetric morphology like some non-thermal SNRs (e.g., SN1006; \cite{Koyama95}).
Figure \ref{fig:flux} right shows a histogram of the calculated surface brightnesses. 
The horizontal arrow represents the 1-$\sigma$ spatial fluctuation of the CXB surface brightness estimated by \citet{Tawa08phd}.
Surface brightnesses in almost all regions ($\sim$94\%) are within the 3-$\sigma$ fluctuation, which suggests that the observed hard X-ray emission is dominated by the CXB.
Although \citet{Katsuda09} argued that non-thermal emission was also detected at the SNR's center, figure \ref{fig:flux} shows that the hard surface brightness from this region is also consistent with CXB fluctuations, as is the case in the NW\_rim.
As a result, there is no robust evidence for non-thermal emission from this remnant.

\subsection{Estimating the age of G156.2+5.7}
As shown above, the observed hard X-ray emission is dominated by the CXB everywhere except in the region of 4$\arcmin$.0 around 2XMM J045637.2+522411.
This fact supports the idea that G156.2+5.7 is an evolved SNR, despite the claim of  \citet{Gerardy07} that it was a young nearby one.
As explained in section \ref{sec:ejecta}, G156.2+5.7 has a layered metallicity structure that is observed in other evolved SNRs (e.g., Cygnus Loop; \cite{Tsunemi07}).  
Such structure suggests that the reverse shock has already reached at the center of G156.2+5.7.
Furthermore, the electron temperature ($<1$ keV) is much lower than that in many young SNRs.
All the results support the picture of an evolved SNR.
For these reasons, we prefer $\sim$1.1 kpc \citep{Pfeffermann91} to $\sim$0.3 kpc \citep{Gerardy07} for its distance $d$ in the following analysis.

Assuming thermal equilibrium between electrons and protons, we find that the forward shock velocity $v_s$ is estimated to be $v_s \simeq614\,(kT_e/0.45 \rm{keV})^{0.5}$ km s$^{-1}$.
However, \citet{Ghavamian01} calculate an initial equilibration of only 50\% for shock speeds of $\sim600$km s$^{-1}$.
In such a non-equilbrium case, the electron temperature $kT_e$, and the proton temperature, $kT_p$ are expressed in the following equations \citep{Ghavamian01}:
\begin{eqnarray}
kT_p \simeq \frac{1}{2}\frac{3}{16} \left[m_e f_{\rm{eq}}+(2-f_{\rm{eq}})m_p\right]v_s^2,\\
kT_e \simeq \frac{1}{2}\frac{3}{16} \left[m_p f_{\rm{eq}}+(2-f_{\rm{eq}})m_e\right]v_s^2,\\
\frac{T_e}{T_p} \simeq \frac{f_{\rm{eq}}}{2-f_{\rm{eq}}},
\end{eqnarray}
where $m_{p,e}$ are the proton and electron masses, respectively.
If the initial electron-proton equilibration, $f_{\rm{eq}}$ is 0.5,  the forward shock velocity $v_s$ is $960\,(kT_e/0.45 \rm{keV})^{0.5}$ km s$^{-1}$, which gives an upper limit of $v_s$.

On the other hand, the ambient densities ($n_0$) for the NW\_rim and the E\_rim are $n_{0,\rm{NW}}=0.084(d/1.1\,\rm{kpc})^{-1}\,\rm{cm^{-3}}$ and $n_{0,\rm{E}}=0.091(d/1.1\,\rm{kpc})^{-1}\,\rm{cm^{-3}}$, respectively, by fitting their emission measure profiles with a Sedov model \citep{Katsuda09}.
Due to the lack of statistics, it is hard to estimate the ambient density for the S\_rim in the same way.
We therefore consider the average value of two calculated $n_0$ to be a global ambient density of G156.2+5.7: hence $n_0=0.088(d/1.1\,\rm{kpc})^{-1}\,\rm{cm^{-3}}$.
Applying a simple Sedov analysis, the age $t_4=t/10^4$ yr is as follows:
\begin{eqnarray}
t_4 \simeq {1.5} \left(\frac{v_s}{614\,\rm{km\,s^{-1}}}\right)^{-5/3}\left(\frac{{\rm E_{0}}}{\rm 10^{51}\,erg}\right)^{1/3}\left(\frac{n_0}{\rm0.088\,cm^{-3}}\right)^{-1/3},
\end{eqnarray}
for the thermal equilibrium case, and
\begin{eqnarray}
t_4 \simeq {0.7} \left(\frac{v_s}{960\,\rm{km\,s^{-1}}}\right)^{-5/3}\left(\frac{{\rm E_{0}}}{\rm 10^{51}\,erg}\right)^{1/3}\left(\frac{n_0}{\rm0.088\,cm^{-3}}\right)^{-1/3},
\end{eqnarray}
for the non-equilibrium case. Here E$_0$ is the explosion energy.
As a result, we prefer the estimate from the ASCA observation (15000 yr; \cite{Yamauchi99}) to that from the ROSAT observation (26000 yr; \cite{Pfeffermann91}).
In any cases, G156.2+5.7 is not a young SNR, as claimed by \citet{Gerardy07}.
While \citet{Katsuda09} pointed out that non-thermal X-ray emission would be detectable even from such an evolved SNR if the flux ratio of non-thermal to thermal emission is higher than those in a typical evolved SNR, it is more reasonable to conclude that G156.2+5.7 is, itself, a typical evolved SNR, since CXB exceeds that of any  X-ray synchrotron component.

\section{Summary}
We performed a set of 6 pointing observations of G156.2+5.7 with Suzaku.
The spectra were well fitted with two-component or single-component VNEI (with CXB and LHB) models for the inner regions and rim regions, respectively.  Our results are consistent with the conclusion of  \citet{Katsuda09}, that these components represent reverse shock-heated ejecta and forward shock-heated ISM, respectively.

From the ejecta component, we found prominent central concentrations of Si, S, and Fe, while the lighter elements (O, Ne, and Mg) are distributed more uniformly.
Such distributions reflect a layered-metallicity structure of the progenitor star.

A single-component VNEI model provided an acceptable fit for all the rim regions; no additional power law component was required for our fit.
We found no robust evidence for the non-thermal emission in the NW\_rim when we excluded the region contaminated by the cluster of galaxies, 2XMM J045637.2+522411.
Accordingly, we conclude that the hard X-ray emission in G156.2+5.7 is sufficiently explained as CXB emission.

The estimated forward shock velocity is relatively slow compared with young SNRs: $v_s \simeq614-960\,(kT_e/0.45 \rm{keV})^{0.5}$ km s$^{-1}$.
We estimated its age to be 7,000-15,000$(d/1.1\,\rm{kpc})^{-1}$ yr; therefore, G156.2+5.7 is not a young, but an evolved SNR.

\section*{Acknowledgments}
H.U. is supported by Japan Society for the Promotion of Science (JSPS) Research Fellowship for Young Scientists. 
H.Y. is supported by JSPS Research Fellowship for Research Abroad.
The work is partially supported by the Ministry of Education, Culture, Sports, Science and Technology (Grant-in-Aid No.23000004).

\end{document}